\documentclass[usenatbib,useAMS,letterpaper]{mnras}
\usepackage{float}
\usepackage{graphicx}
\usepackage{epstopdf}
\usepackage{ textcomp }
\usepackage{multirow}
\usepackage{ctable}
\usepackage{url}
\usepackage{times}
\usepackage{amsmath}
\usepackage{amssymb}
\usepackage{xspace}
\usepackage{mathrsfs} 
\usepackage{tabularx}
\usepackage{fixltx2e}
\usepackage{color}
\usepackage{placeins}
\usepackage{comment}

\newcommand{\acknowledgments}{\begin{small}\section*{Acknowledgements}\end{small}}

\defcitealias{2019MNRAS.487.4393S}{Paper {\small I}}
\defcitealias{2020MNRAS.491.1190S}{Paper {\small II}}
\newcommand\sref[1]{\hyperref[#1]{\S~\ref*{#1}}}
\newcommand\fref[1]{\hyperref[#1]{Fig.~\ref*{#1}}}
\newcommand\Eqref[1]{Eq.~(\hyperref[#1]{\ref*{#1}})}
\newcommand\eeqref[1]{Eq.~\hyperref[#1]{\ref*{#1}}}

\newcommand\tref[1]{\hyperref[#1]{Table~\ref*{#1}}}
\newcommand\aref[1]{\hyperref[#1]{Appendix~\ref*{#1}}}

\defcitealias{2019MNRAS.487.4393S}{Paper {\small I}}
\defcitealias{2020MNRAS.491.1190S}{Paper {\small II}}

\newcommand{\oneline}[1]{%
  \newdimen{\namewidth}%
  \setlength{\namewidth}{\widthof{#1}}%
  \ifthenelse{\lengthtest{\namewidth < \textwidth}}%
  {#1}
  {\resizebox{\textwidth}{!}{#1}}
}
 
\title[Satellites Seeding the CGM with Cold gas]{Seeding the CGM: How Satellites Populate the Cold Phase of Milky Way Halos}
\author[Roy et al.]{
\parbox[t]{\textwidth}{
Manami Roy$^{1,2}$ \thanks{E-mail: roy.516@osu.edu}, Kung-Yi Su$^{3}$, Stephanie Tonnesen$^{4}$,
Drummond B. Fielding$^{4}$, Claude-André Faucher-Giguère$^{5}$
\\ \\
\footnotesize{$^{1}$ \textit{Raman Research Institute, 
Sadashiva Nagar, 
Bangalore 560080, India} \\
$^{2}$\textit{Center for Cosmology and Astro Particle Physics (CCAPP), The Ohio State University, 191 W. Woodruff Avenue, Columbus, OH 43210, USA} \\
$^{3}$ \textit{Black Hole Initiative, Harvard University, 20 Garden St., Cambridge, MA 02138, USA} \\
$^{4}$ \textit{Center for Computational Astrophysics, Flatiron Institute, 162 5th Ave, New York, NY 10010, USA}\\
$^{5}$ \textit{Department of Physics and Astronomy and CIERA, Northwestern University, 2145 Sheridan Road, Evanston, IL 60208, USA} 
}}
\vspace*{6pt} \\
}
\usepackage[all]{hypcap}

\begin{document}
\long\def\/*#1*/{}
\date{Accepted for publication in MNRAS on Oct 6, 2023}

\pagerange{\pageref{firstpage}--\pageref{lastpage}} \pubyear{2023}

\maketitle

\label{firstpage}

\begin{abstract}
The origin of the cold phase in the CGM is a highly debated question. We investigate the contribution of satellite galaxies to the cold gas budget in the circumgalactic medium (CGM) of a Milky Way-like host galaxy. We perform controlled experiments with three different satellite mass distributions and identify several mechanisms by which satellites can add cold gas to the CGM, including ram pressure stripping and induced cooling in the mixing layer of the stripped cold gas. These two mechanisms contribute a comparable amount of cold gas to the host CGM. We find that the less massive satellites ($\leq 10^9 M_\odot$) not only lose all of their cold gas in a short period ($\sim$ 0.5-1 Gyr), but their stripped cold clouds also mix with the hot CGM gas and get heated up quickly. However, stellar feedback from these less massive satellites can hugely alter the fate of their stripped gas. Feedback speeds up the destruction of the stripped cold clouds from these satellites by making them more diffuse with more surface area. On the other hand, the more massive satellites (LMC or SMC-like $\sim 10^{10} M_\odot$) can add cold gas to the total gas budget of the host CGM for several Gyrs. 
\end{abstract}

\begin{keywords}
methods: numerical --- galaxies: evolution --- Galaxy: halo
\end{keywords}

\section{Introduction}
\label{S:intro}
A significant portion of galactic baryonic content resides in the form of a diffuse gaseous halo, known as the Circumgalactic medium (CGM), which surrounds the galactic disk and extends up to the virial radius and even beyond \citep[for a comprehensive review see][]{Tumlinson2017,Faucher2023}. 

Absorption and emission observations of the CGM make it abundantly clear that the gas in the CGM is multiphase in nature. On the basis of their temperatures, these phases are roughly divided into hot (T $>10^6$K), warm ($10^5-10^6$K) and cold phases ($<10^4$K). Recent observations in massive halos ($10^{11-13} \text{M}_\odot$) showed high column densities of MgII and HI, which are the tracers of the cold phase of CGM  \citep{Zhu2014,Chen2018,Zahedy2018}, even out to the virial radius. The observations by \cite{LanMo2018,LanMo2019} also indicated the existence of a cold phase out to large radii ($>100$kpc). This leads to two highly debated questions:  \textit{how do these massive halos whose virial temperature is much higher than that of the cold phase form cold gas?} and \textit{how does the cold gas exist at such large radii?} 
\begin{figure*}
\includegraphics[width=1.0\textwidth]{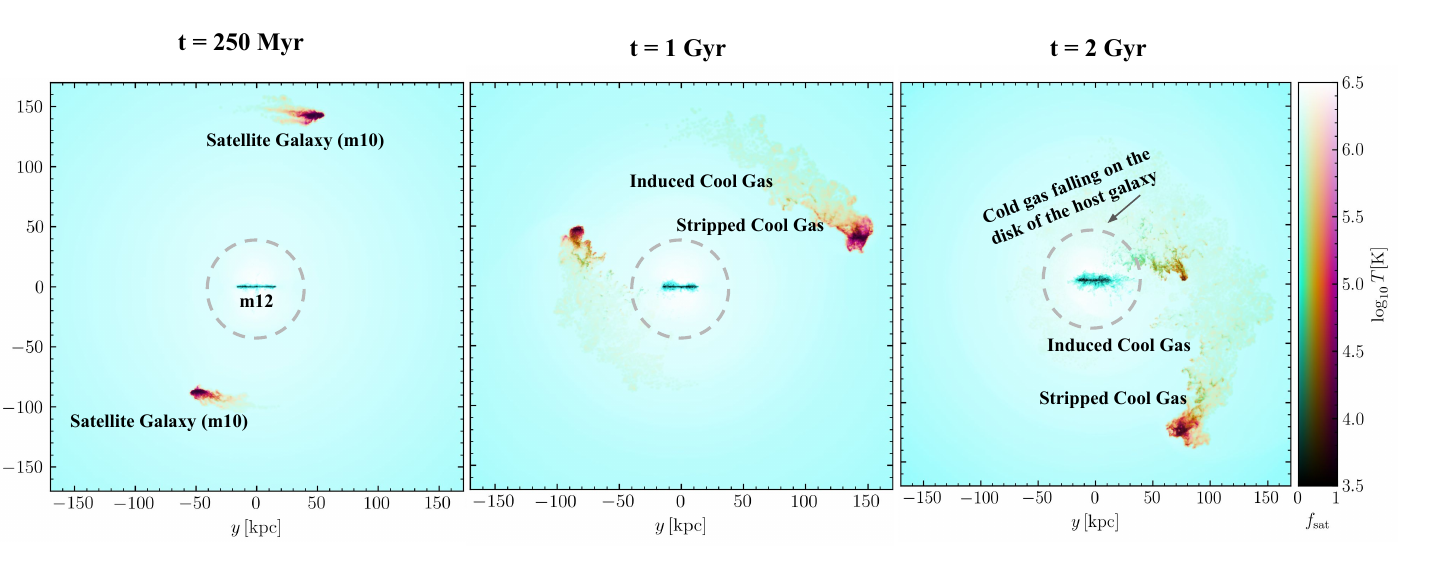}
\caption{{The temperature distribution of three snapshots in simulation (2xm10\_far\_highres, see Table \ref{runs}). The colorbar varies both color and saturation based on $f_{\rm sat}$ and temperature, respectively. The parameter $f_{\rm sat}$ represents the local mass fraction that originated from the satellite such that $f_{\rm sat}=1$ is entirely composed of satellite gas and $f_{\rm sat}=0$ is entirely host gas. The stripped cold gas is streaming behind the satellites and falling towards the central disk. There is also induced cool gas in the mixing layer of stripped cool gas and hot host gas.}}
\label{sch_m10}
\end{figure*} 

\begin{figure*}
\includegraphics[width=1.0\textwidth]{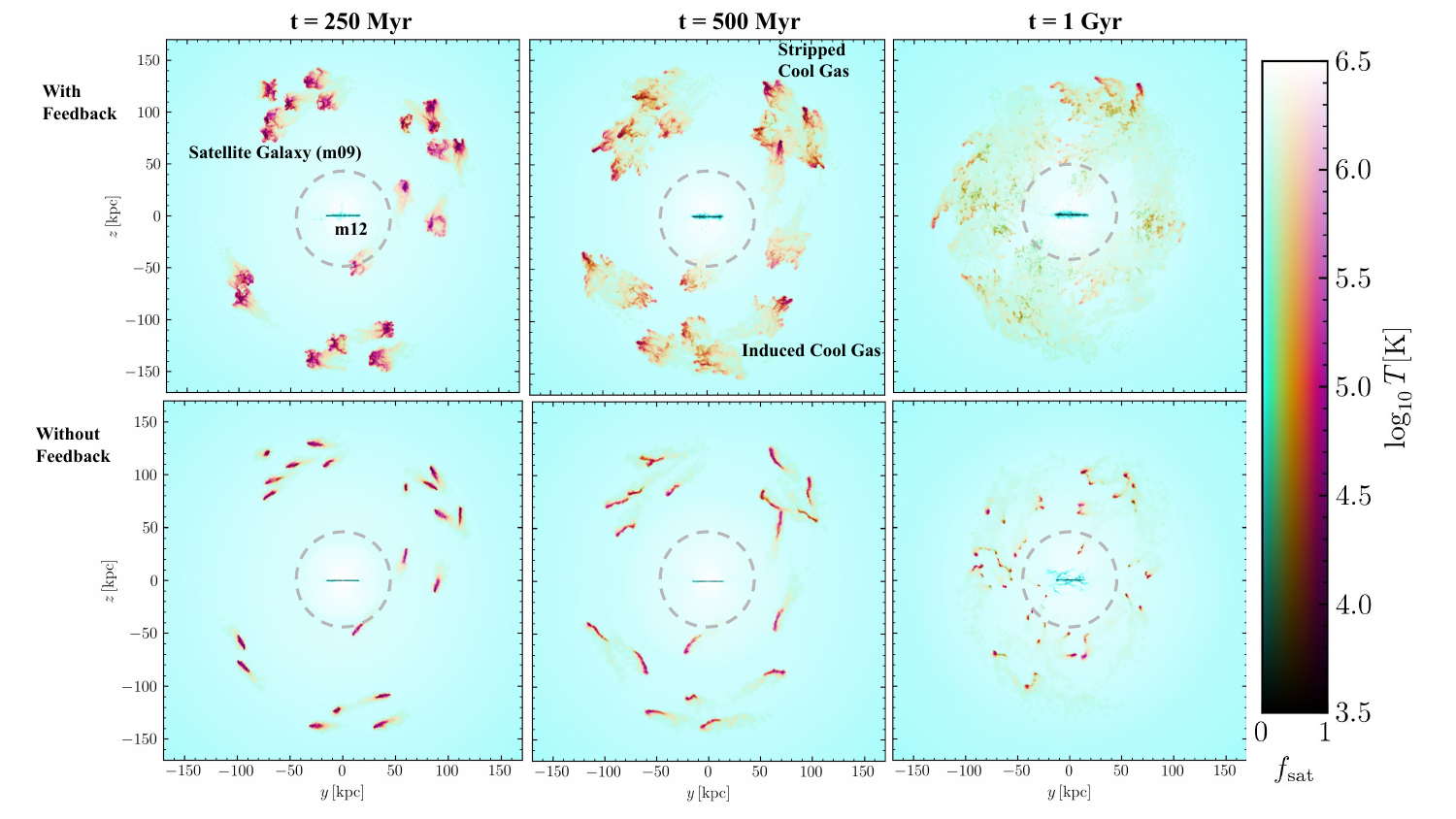}
\caption{{The temperature distribution of the three snapshots in simulations (20xm09\_far\_highres and 20xm09\_far\_highres\_noFB, see Table \ref{runs}), with the upper and lower panels showing the runs with and without the inclusion of feedback, respectively. The colorbar is as described in Figure \ref{sch_m10}.
Each orange galaxy tail is shorter than for the m10 satellites, but because there are more satellites the high $f_{\rm sat}$ gas covers more area.}}
\label{sch_m09}
\end{figure*} 

\begin{figure*}
\includegraphics[width=1.0\textwidth]{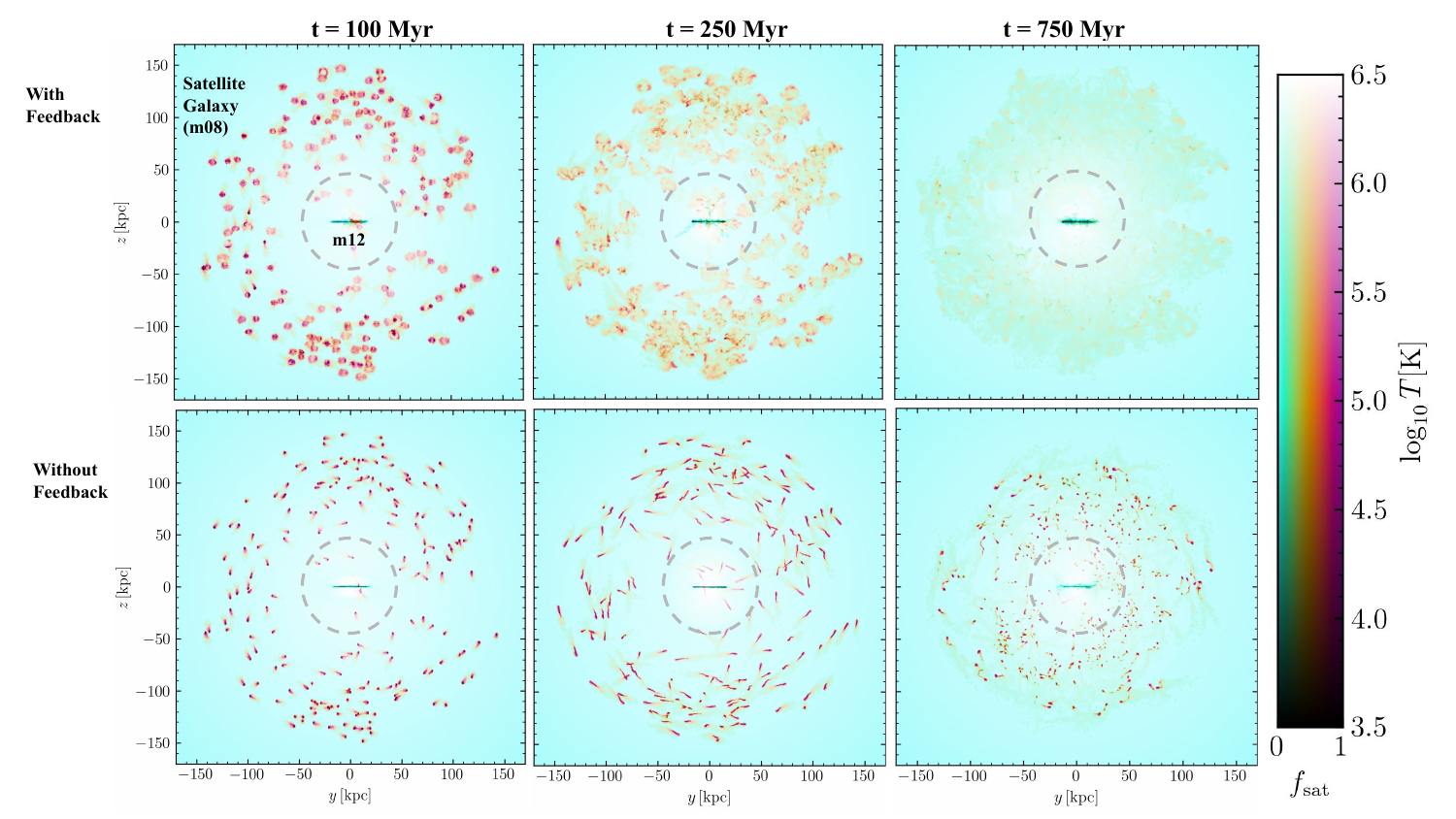}
\caption{{The temperature distribution of the three snapshots in simulations (200xm08\_far\_highres and 200xm08\_far\_highres\_noFB, see Table \ref{runs}), with the upper and lower panels showing the runs with and without the inclusion of feedback, respectively. The colorbar is as described in Figure \ref{sch_m10}.  The impact of feedback is clear in the more diffuse distribution of satellite gas in the upper panels.}}
\label{sch_m08}
\end{figure*} 

\begin{table*}
\begin{center}
 \caption{Properties of Initial Conditions for the Simulations/Halos Studied Here}
 \label{tab:ic}
 \begin{tabular*}{\textwidth}{@{\extracolsep{\fill}}l|cc|ccc|c|cc|cc|cc|ccc}
 \hline
\hline
&\multicolumn{2}{c|}{\underline{Resolution}}&\multicolumn{3}{c|}{\underline{DM halo}}&&\multicolumn{2}{c|}{\underline{Stellar Bulge}}&\multicolumn{2}{c|}{\underline{Stellar Disc}}&\multicolumn{2}{c|}{\underline{Gas Disc}}&\multicolumn{2}{c}{\underline{Gas Halo}}  \\
$\,\,\,\,$Model  &$\epsilon_g$ &$m_g$ (hr/lr)        &$M_{\rm halo}$   &$r_{dh}$            &$V_{\rm Max}$    &$M_{\rm bar}$    &$M_b$ 
                 &$a$          &$M_d$        & $r_d$             &$M_{gd}$       &$r_{gd}$         &$M_{gh}$         &$r_{gh}$     \\
                 &(pc)         &(M$_\odot$)  &(M$_\odot$)      & (kpc)           &(km/s)           &(M$_\odot$)      &(M$_\odot$) 
                  &(kpc)        &(M$_\odot$)  &(kpc)            &(M$_\odot$)    &(kpc)            &(M$_\odot$)      &(kpc)\\
\hline
\multicolumn{15}{c}{{\bf Host galaxy}}\\
\hline
$\,\,\,\,$m12        &1       &8e3 / 8e4           &1.8e12           &20             &174              &3.2e11           &1.5e10   
                     &1.0       &5.0e10          &3.0                &5.0e9            &6.0                &2.5e11           &20        \\  
\hline
\multicolumn{15}{c}{{\bf Satellite galaxy}}\\
\hline
$\,\,\,\,$m10        &1    &8e3 / 8e4           &2e10            &4.7             &35.2              &7.3e8           &1e7   
                     &1.5       &3.0e8          &0.7                &4.2e8            &2.1                & -           & -        \\                
$\,\,\,\,$m09        &1     &8e3 / 4e4           &2e9            &2.2             &16.4              &7.2e7           &2e6   
                     &0.223       & -          & -               &7e7            &0.87                & -           & -        \\ 
$\,\,\,\,$m08        &1     &8e3 / 4e4           &2e8            &0.9             &7.62              &7.3e6           &8e4   
                     &0.045       & -          & -               &7.2e6            &0.27                & -           & -        \\                      

\hline 
\hline
\end{tabular*}
\end{center}
\begin{flushleft}
Parameters of the galaxy models studied here : 
(1) Model name. The number following `m' labels the approximate logarithmic halo mass. 
(2) $\epsilon_g$: Minimum gravitational force softening for gas (the softening for gas in all simulations is adaptive, and matched to the hydrodynamic resolution; here, we quote the minimum Plummer equivalent softening).
(3) $m_g$: Gas mass (resolution element). There is a resolution gradient for m14, so its $m_g$ is the mass of the highest resolution elements.
(4) $M_{\rm halo}$: Dark matter halo mass within $R_{\rm vir}$. 
(5) $r_{dh}$: NFW halo scale radius (the corresponding concentration of m12,m13,m14 is $c=12,\,6,\,5.5$).
(6) $V_{\rm max}$: Halo maximum circular velocity.
(7) $M_{\rm bar}$: Total baryonic mass within $R_{\rm vir}$. 
(8) $M_b$: Bulge mass.
(9) $a$: Bulge Hernquist-profile scale-length.
(10) $M_d$ : Stellar disc mass.
(11) $r_d$ : Stellar disc exponential scale-length.
(12) $M_{gd}$: Gas disc mass. 
(13) $r_{gd}$: Gas disc exponential scale-length.
(14) $M_{gh}$: Hydrostatic gas halo mass within $R_{\rm vir}$. 
(15) $r_{gh}$: Hydrostatic gas halo $\beta=1/2$ profile scale-length.
\end{flushleft}
\end{table*}

\vspace{-0.2cm}

\begin{table*}
	\caption{Summary of all the runs}
	\begin{tabular}{||c|c|c|c|c|c|c||}
 \hline
 
Model & Satellite Number & Satellite Position & Resolution & Feedback & Gas & Symbol \\ [0.7ex] 
 \hline\hline
 \multirow{3}{*}{m10} 
 & 2 & One at 100 kpc and one at 150 kpc  & lr & Yes & Yes & 2xm10\_far\_
lowres \\
 & 2 & One at 100 kpc and one at 150 kpc & hr & Yes & Yes & 2xm10\_far\_highres \\ 
& 2 & One at 50 kpc and one at 100 kpc & lr & Yes & Yes & 2xm10\_near\_lowres \\ 
 \hline \hline
 \multirow{7}{*}{m09} 
& 20 & Randomly placed between 100 kpc and 150 kpc  & lr & Yes & Yes & 20xm09\_far\_
lowres \\
& 20 & Randomly placed between 100 kpc and 150 kpc & hr & Yes & Yes & 20xm09\_far\_highres \\ 
& 20 & Randomly placed between 50 kpc and 150 kpc & lr & Yes & Yes & 20xm09\_near\_lowres \\ 
& 10 & Randomly placed between 100 kpc and 150 kpc  & lr & Yes & Yes & 10xm09\_far\_
lowres \\
& 40 & Randomly placed between 100 kpc and 150 kpc & lr & Yes & Yes & 40xm09\_far\_
lowres \\ 
& 20 & Randomly placed between 100 kpc and 150 kpc & lr & No & Yes & 20xm09\_far\_
lowres\_noFB \\ 
& 20 & Randomly placed between 100 kpc and 150 kpc & lr & Yes & No & 20xm09\_far\_
lowres\_nogas \\ 
 \hline \hline 
 \multirow{4}{*}{m08} 
& 200 & Randomly placed between 100 kpc and 150 kpc  & lr & Yes & Yes & 200xm08\_far\_
lowres \\
& 200 & Randomly placed between 100 kpc and 150 kpc & hr & Yes & Yes & 200xm08\_far\_highres\\ 
& 200 & Randomly placed between 50 kpc and 150 kpc & lr & Yes & Yes & 200xm08\_near\_lowres \\ 
& 200 & Randomly placed between 100 kpc and 150 kpc & hr & No & Yes & 200xm08\_far\_highres\_noFB \\ 
 \hline
 \end{tabular}
 \vspace{1ex}

\label{runs}     
\end{table*}
Recent studies have shown that thermally unstable perturbations can drive cooling in the CGM, proceeding into multiphase condensation, if the ratio of the radiative cooling time to free-fall time falls below a threshold value \citep{MB2004,McCourt2011,Sharma2012,Voit2015,Fielding2017,Voit2019}. In contrast, \cite{Esmerian2021} showed that this threshold is a poor predictor of whether the density perturbation in hot CGM gas leads to cooling or not. Star-formation-driven outflows can also uplift cold gas to the CGM from the galactic disc \citep{Faucher2015,Faucher2016,Liang2016}. However, the time taken for cold clouds to reach $>100$kpc distance from the disc is greater than $10$ Myr, which is much larger than the cloud crushing time for a typical $100$ pc cloud ($<1$ Myr; See Equation \ref{tcc}) \citep{Pro2013,Zhang:2017}. Therefore, it is challenging for star-formation-driven outflows to populate the outer CGM with cold gas. However, galaxies can accrete fresh cold gas directly from cold dense filaments of Intergalactic medium (IGM), known as cold mode accretion \citep{Birnboim2003,Keres2005,FG2011,van2011}. Along with this cold mode accretion, satellite galaxies can also populate the outer, as well as inner, CGM with cold gas \citep{Suresh2019}.

A recent study \citep{Fielding2020} has compared the results from different idealized and cosmological simulations and concluded that more cold gas in the outer CGM was found (see their Figure 3) in cosmological simulations \citep{Joung2012,Mari2018,Nelson2018} than in isolated galaxy simulations \citep{Fielding2017,Li2020,Su2020}, especially at large radii (beyond  $\sim$0.5\ r/r$_{200}$). The idealized simulations did not include either cold mode IGM accretion or satellite galaxies, either of which could be responsible for adding cold gas to the outer CGM in cosmological simulations.
 However, in cosmological simulations, it is challenging to distinguish the amount of cold phase in the CGM contributed only by satellite galaxies from feedback-driven cold clouds or cold filamentary inflows. In this paper, we run a suite of high-resolution idealized simulations of Milky Way-type host galaxies, varying the mass and spatial distribution of satellite galaxies in each run. This will allow us to explicitly determine the amount and processes by which satellites can populate the cold-phase of the CGM of their host galaxy. 

When a satellite galaxy passes through the diffuse gas of the CGM, it experiences a headwind that causes pressure on the galaxy, known as `Ram Pressure'. Its magnitude depends on the relative speed of the satellite with respect to the medium and the local density of the medium. If this ram pressure exceeds the local gravitational restoring pressure of the satellite galaxy, its gas can be stripped \citep{Gunn1972}. This is known as `Ram Pressure Stripping'. In lower-mass galaxies, ram pressure stripping becomes an effective mechanism for removing gas due to their lower gravitational restoring force \citep{Samuel2022,Saee2023}. Also, galaxies moving through the CGM of massive halos will experience higher ram pressure due to a combination of higher CGM density and faster orbital velocities. Ram pressure stripping can be an important factor in the quenching of  satellite galaxies by removing their fuel for star formation \citep{Samuel2022_q}.

There is significant observational evidence that ram pressure not only removes gas from satellite galaxies, but it also populates the CGM of the host galaxies with cold gas. For example, the neighboring dwarf galaxies of the Milky Way and M31 system tend to be poorer in HI gas content than those at larger distances \citep{Gr2009,Putman2021}. In addition to this, it is also apparent from recent MUSE observations that there is a strong connection between the group environment and the ionization structure of the CGM. These observations showed that moving from lower mass systems to group environments leads to a significant increase in the covering fractions of MgII and HI ions at fixed impact parameter (in kpc) of the CGM of host galaxies \citep{ Dutta2021}. Therefore, the group environment contributes more cold gas in the galaxies than the isolated systems. Along with this, cold gas can be stripped from the satellites due to the ejecting wind, and can be found behind the satellites in the form of a wake 
\citep{Ostriker1999,Bernal2013}. 
The cold gas, which directly gets stripped from the satellites, will mix with the hot CGM, and in the mixing layer of this stripped cold gas significant cooling can occur \citep{Tonnesen2021}. The satellites can also stir the CGM gas and create local perturbation which can lead to the condensation of cold gas out of hot CGM gas \citep{Sharma2012,McCourt2011,Voit2018,Esmerian2021}. The relative importance of these different mechanisms is still poorly constrained. 
In this paper, we will separate these different mechanisms and investigate the amount of cold gas contributed by each mechanism. 

The paper is structured as follows. In Section \ref{S:methods}, we discuss the methodology of our simulation, where we describe the initial conditions of our simulation setup (Section \ref{S:ic}) along with our definition of cold CGM gas for our analysis (Section \ref{S:Define Cold Gas}). In Section \ref{S:results}, we demonstrate our results from our analysis of the simulations, where we discuss the origin (Section \ref{S:origin}) and amount of the cold gas (Section \ref{S:amount}) in the CGM. In addition, we also describe how different satellite properties such as satellite mass distribution (Section \ref{S:mass dist} and \ref{S:survival}), spatial location (Section \ref{S:location}), stellar feedback (Section \ref{S:feedback}), number of satellites (Section \ref{S:number}), and resolution of the simulation (Section \ref{S:resolution}) affect the cold gas mass in the CGM.
In Section \ref{S:turb}, we discuss the amount of turbulence driven cooling in the CGM. In Section \ref{S:obs}, we demonstrate how our estimated cold gas mass lines up with observational values followed by a discussion about other sources of cooling in the CGM in Section \ref{S:other cooling}. Finally, we summarize our results and discuss future work in Section \ref{S:conclusion}.

\section{Methodology} \label{S:methods}
Our simulations use {\sc GIZMO}\footnote{A public version of this code is available at \href{http://www.tapir.caltech.edu/~phopkins/Site/GIZMO.html}{\textit{http://www.tapir.caltech.edu/$\sim$phopkins/Site/GIZMO.html}}}  \citep{2015MNRAS.450...53H}, in its meshless finite mass (MFM) mode, which is a Lagrangian mesh-free Godunov method, capturing the advantages of grid-based and smoothed-particle hydrodynamics (SPH) methods. Numerical implementation details and extensive tests are presented in a series of methods papers for, e.g.,\ hydrodynamics and self-gravity \citep{2015MNRAS.450...53H}, magnetohydrodynamics \citep[MHD;][]{2016MNRAS.455...51H,2015arXiv150907877H}, anisotropic conduction and viscosity \citep{2017MNRAS.466.3387H,2017MNRAS.471..144S}, and cosmic rays \citep{chan:2018.cosmicray.fire.gammaray}.

All of our simulations except for the runs with no feedback, have the FIRE-2 implementation of the Feedback In Realistic Environments (FIRE\footnote{\href{http://fire.northwestern.edu}{\textit{http://fire.northwestern.edu}}}) physical treatments of the ISM, star formation, and stellar feedback, the details of which are given in \citet{hopkins:sne.methods,2017arXiv170206148H} along with extensive numerical tests.  Cooling is followed from $10-10^{10}$K, including the effects of photo-electric and photo-ionization heating, collisional, Compton, fine structure, recombination, atomic, and molecular cooling. 

Star formation is treated via a sink particle method, allowed only in molecular, self-shielding, locally self-gravitating gas above a density $n>100\,{\rm cm^{-3}}$ \citep{2013MNRAS.432.2647H}. Star particles, once formed, are treated as a single stellar population with metallicity inherited from their parent gas particle at formation. All feedback rates (SNe and mass-loss rates, spectra, etc.) and strengths are IMF-averaged values calculated from {\small STARBURST99} \citep{1999ApJS..123....3L} with a \citet{2002Sci...295...82K} IMF. The stellar feedback model includes (1) Radiative feedback, including photo-ionization and photo-electric heating, as well as single and multiple-scattering radiation pressure tracked in five bands  (ionizing, FUV, NUV, optical-NIR, IR), (2) OB and AGB winds, resulting in continuous stellar mass loss and injection of mass,  metals, energy, and momentum  (3) Type II and Ia SNe (including both prompt and delayed populations) occurring according to tabulated rates and injecting the appropriate mass, metals, momentum, and energy to the surrounding gas. All the simulations  also include MHD, fully anisotropic conduction, and viscosity with the Spitzer-Braginski coefficients.

\begin{figure}
\includegraphics[width=1.0\linewidth]{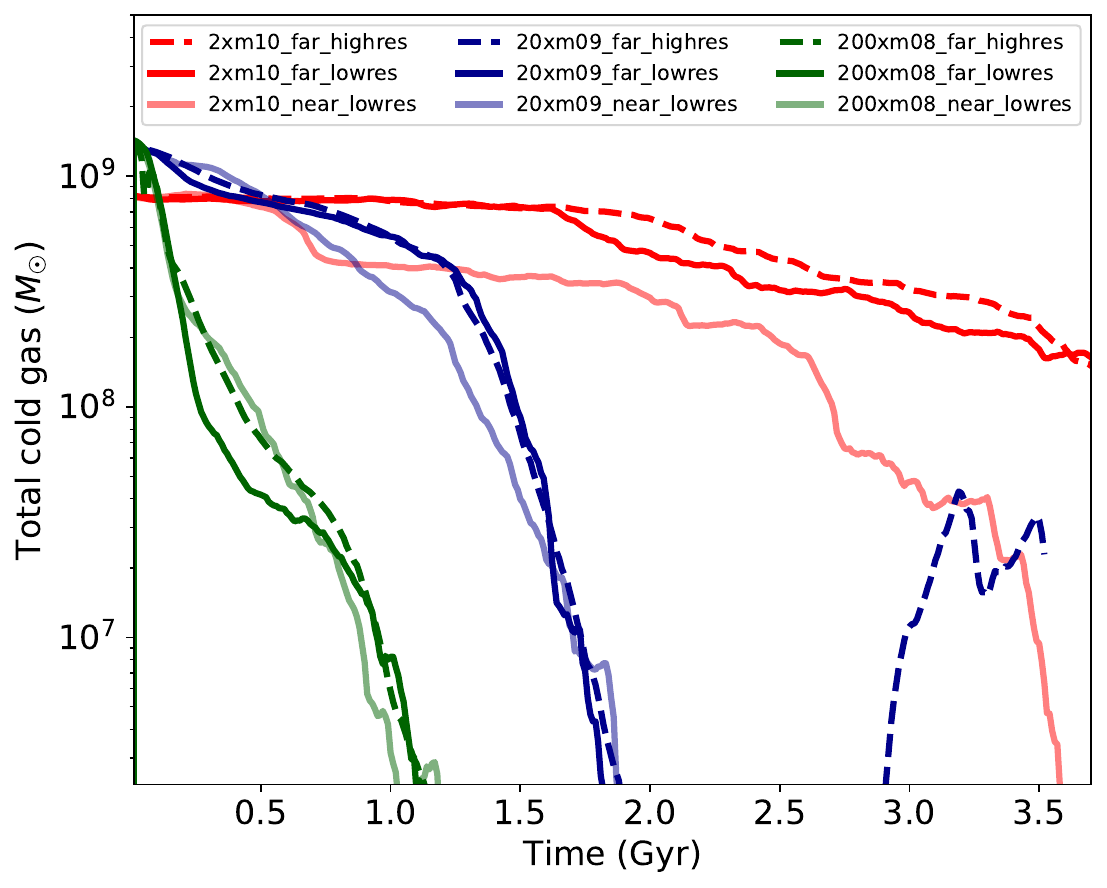}
\caption{{The time evolution of total cold (T$\le3\times10^4$K) gas mass beyond $40$kpc radius from the center of the host galaxy. This total cold gas mass includes cold gas inside the satellites, cold gas stripped from the satellites, and cold gas in the host CGM. Labels for the runs are described in Table \ref{runs}.}}
\label{coldgasmass_tot}
\end{figure} 

\begin{figure}
\includegraphics[width=8cm,height=21cm]{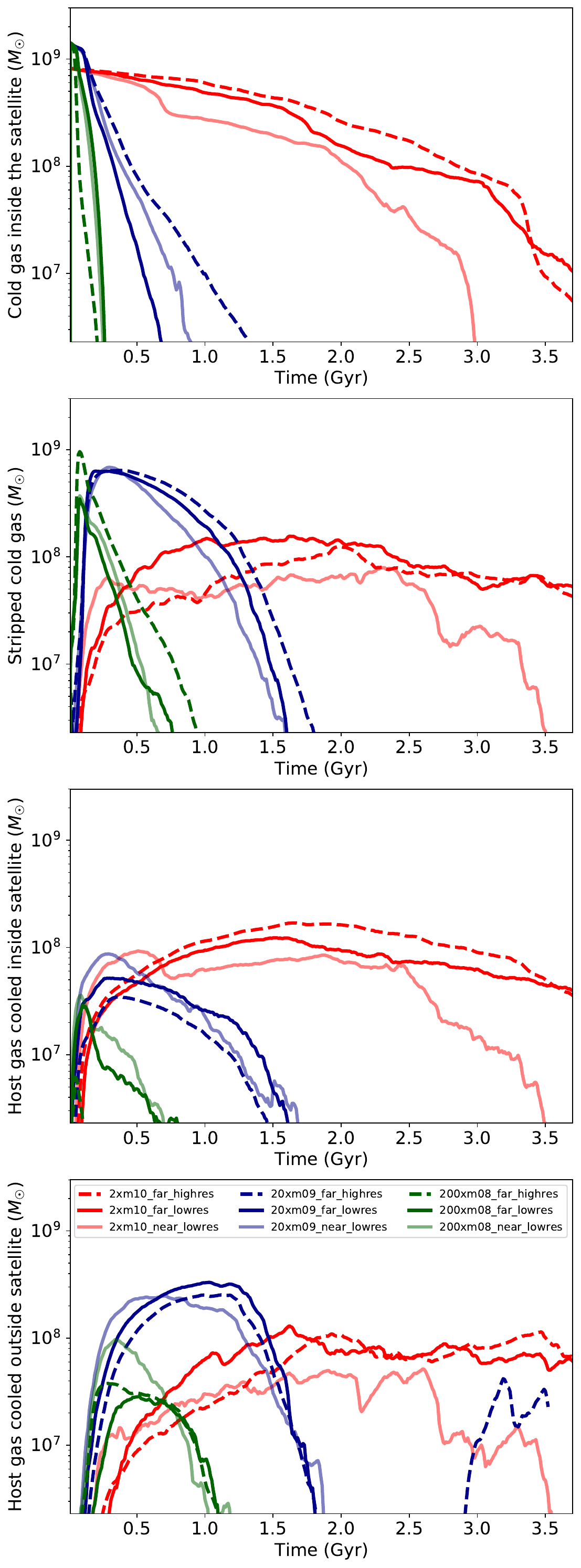}
\caption{{The time evolution of the cold (T$\leq3\times10^4$K) gas mass from different contributions beyond $40$kpc radius from the center of the host galaxy. From top to bottom, the panels respectively describe cold gas inside the satellites, cold gas stripped from the satellites (defined as cold satellite gas which lies beyond 6 times scale radius of the satellites), host cold gas cooled inside of the satellites, and host gas cooled outside of the satellites.}}
\label{coldgasmass1}
\end{figure} 

\subsection{Initial Conditions}
\label{S:ic}

The initial conditions studied here mostly follow what is described in detail in \cite{2019MNRAS.487.4393S}. To further stabilize the host CGM, we expand the simulation region to 3 times the viral radius, and the simulations are run adiabatically (no cooling or star formation) for 4.5 Gyr to relax any initial transients before the satellites are placed into the CGM. 
The simulation properties are summarized in \tref{tab:ic}. In this paper, our study will focus on the {\bf m12} halo. The dark matter (DM) halo, bulge, black hole, and gas+stellar disk are initialized following  \cite{1999MNRAS.307..162S,2000MNRAS.312..859S}.
We assume a spherical, isotropic, \citet{1996ApJ...462..563N} profile DM halo; a \cite{1990ApJ...356..359H} profile stellar bulge; an exponential, rotation-supported disk of gas and stars ($10^{9}$ and $5\times10^{10} M_\odot$, respectively) initialized with Toomre $Q\approx1$; a BH with mass $\sim1/300$ of the bulge mass \citep[e.g.][]{2004ApJ...604L..89H}; and an extended spherical, hydrostatic gas halo with a $\beta$-profile ($\beta=1/2$) and rotation with constant angular momentum at $35$ kpc. {Even though the simulations are initiated in hydrostatic equilibrium, they do not remain in hydrostatic equilibrium after the simulations start. They gradually migrate to a cooling flow solution \citep{Stern2019}.} The initial metallicity drops from solar ($Z=0.02$) to $Z=0.001$ with radius as $Z=0.02\,(0.05+0.95/(1+(r/20\,{\rm kpc})^{1.5}))$. Initial magnetic fields are azimuthal with $|{\bf B}|=0.03\,\mu{\rm G}/(1+(r/20\,{\rm kpc})^{0.375})$. {These initial conditions are quite reasonable as discussed in \cite{Su2019}, with the CGM having temperature around viral temperature, reasonable density profile, and migrating toward the cooling flow solution. However, we do not include filamentary cold accretion or cosmological process in the initial conditions, which are left for future work.}

We consider three different distributions of the satellites with different masses ranging from very small satellites of DM halo mass $2\times10^{8}\,M_{\odot}$ to SMC-like satellites of mass $2\times10^{10}\,M_{\odot}$. We balance the satellite mass and number distributions such that the total DM halo mass of the satellites are same for all the cases. The three distributions that we simulate are the following: 1) 2 satellites of $2\times10^{10}\,M_{\odot}$ (m10), 2) 20 satellites of $2\times10^{9}\,M_{\odot}$ (m09), 3) 200 satellites of $2\times10^{8}\,M_{\odot}$ (m08). We also run additional variations to test the impact of the number of satellites, the gas mass in satellites, and stellar feedback.  To that end, we vary the number of satellites in our fiducial run of m09 from 20 to 10 and 40. We also run a fiducial setup (m09) with no gas at all inside the satellites. Finally, to study feedback, we have two fiducial runs (m08 and m09) with no stellar feedback in the host and the satellites. We summarize all of our runs in Table \ref{runs}.

The satellites are initialized with the same method as the host described above except without a CGM gas halo. The properties are also summarized in \tref{tab:ic}. The m10 galaxy properties are set following the SMC in \cite{2010ApJ...721L..97B,2011MNRAS.417..950H,2015arXiv151103346B,2017MNRAS.471..144S}, with most of the stellar mass in the disk. 
For m09 and m08, the dark matter scale radius roughly follows \cite{2020Natur.585...39W}, with the stellar population modeled as a bulge with mass following \cite{2017MNRAS.467.2019R} and size following a constant surface density ($\sim 5 M_\odot {\rm pc}^{-2}$; \citealt{2020MNRAS.495...78S}), and the ISM gas is set so that the baryon fractions ($\sim 3.5\%$) and the gas surface density are the same as m10.

\subsection{Defining Cold CGM Gas}
\label{S:Define Cold Gas}
In this section, we define cold gas in the CGM before going into the question of how much cold gas is contributed by satellites. We consider gas to be cold if it has a temperature of less than $3\times10^4$ K. As we are interested in the contribution to the cold phase of the CGM by satellite galaxies, we exclude the ISM of the host galaxy by excluding all gas within a radius of 40 kpc ($\sim 0.15$ R$_{\rm vir}$) from the center of the host galaxy for our analysis.    

In addition, we distinguish among different origins of the cold gas in our simulation, such as ram-pressure stripping, induced cooling in the mixing layer. The gas surface density ($\Sigma$) of the satellites becomes much lower than the central gas surface density ($\Sigma_0$) at the gas scale radius, $r_{gd}$ ($r_{gd} \sim 2.7\times r_{d}$, where $r_d$ is stellar scale radius) as $\Sigma = \Sigma_0 \times e^{r/r_d}$ \citep{Krav2013}. Hence, we choose 6 times the gas scale radius of the satellite to be the radius beyond which the gravitational pull from the satellites is negligible. 

 With the Lagrangian MFM method in GIZMO, we can track individual gas particle over the duration of simulation. We use them to define each cooling channel from either satellite or CGM gas for the sake of later analysis: 
\begin{itemize}
\item \textbf{Stripped (cold) gas:} Particles that are initially within the satellite, but later move outside of satellite (and is cooler than $3\times10^4$K). Therefore, we consider the satellite gas to be stripped if it moves from within to outside of 6 times the satellite gas scale radius from the center of the satellite. The coordinates of the satellite black hole are considered to be the center of the satellite.
\item \textbf{CGM cooled inside satellite:} Particles initially belonging to the host CGM that later become $\le 3\times10^4$K and reside inside the satellite.
\item \textbf{CGM cooled outside satellite:} Particles initially belonging to the host CGM that later become  $\le 3\times10^4$K and have never been inside satellite; host gas particles that are cooled outside of the satellite (which is defined above as beyond 6 times the satellite gas scale radius)
\end{itemize}

To summarize, we define cold CGM gas by different cuts in temperature and radius of the host. With that definition, we also distinguish their different origin with particle tracking and cut based on the radius of the satellite galaxies. The following section discusses our main findings based on these definitions.
\section{Results} \label{S:results}
The main goals of this paper are to find how much cold gas (T$<3\times10^4$ K) is contributed by satellite galaxies to the CGM of the Milky Way-type host galaxy, by what mechanisms the satellites increase the cold gas mass, and how satellite properties affect the cold gas mass. Here we present our findings, first grounding our intuition with snapshots from the simulations, then discussing quantitative measures of the cold gas mass.

\subsection{Where is the cold gas?} \label{S:origin}
In this section, we will give an overview about what happens to the CGM gas of the host galaxy using snapshots from several simulations. We show snapshots of temperature distribution for three runs of 2xm10\_far\_highres, 20xm09\_far\_highres, and 200xm08\_far\_highres (See Table \ref{runs}) in Figure \ref{sch_m10}, \ref{sch_m09}, and \ref{sch_m08} respectively. 
We smooth the gas particle data into a $1024^3$ regular cells with an SPH-like deposition according to the particle smoothing length. We mass-weighted the temperature in each grid while depositing. We weight the temperature by $n^2$ (roughly luminosity weighted) along the projection. 
 Satellite and host gas are indicated by the parameter `satellite fraction' ($f_{\rm sat}$). For satellite fraction, we mass-weighted along the direction of projection excluding anything from the host and $>3\times10^5$ K from the average. $f_{\rm sat}=1$ indicates purely satellite gas, while $f_{\rm sat}=0$ denotes pure host gas. The temperature is shown via color saturation.  These all are edge on projections relative to the host galaxy of different simulation snapshots. At the center (0,0), the thin blue strip is nothing but the host ISM and around that the circle with $40$ kpc radius denotes our radial cut to exclude the host ISM. To illustrate the effect of feedback, we show in the top and bottom panels of Figure \ref{sch_m09} and \ref{sch_m08} the runs with feedback and without any feedback, respectively.


Although these figures all show snapshots from different times, they follow the same general trends.  At the earliest time, we can see that most of the cold gas is associated with individual satellites with short tails of cold gas streaming behind them.  As we step forward in time, the tails become longer, and the cold gas becomes free from the satellite's gravitational pull and falls towards the central galaxy.  In addition, the gas becomes more mixed with the host CGM, illustrated by the color change from red to orange. 
 Importantly, we note that cold gas is found either near the central disk or supernova-driven outflows or near the stripped tails of gas.  As we move along stripped tails of gas the $f_{\rm sat}$ value smoothly decreases, indicating mixing. Even where the color indicates that the gas largely originated in the host halo, there is a clear spatial correlation with stripped tails.
 
 
 We can use this visual inspection as a first step in determining how host CGM gas cools in these simulations.  There are two likely ways this can happen: satellites can stir the CGM of the host galaxy and create local thermal instabilities which can subsequently lead to cooling (turbulence-driven cooling). In addition, cold-stripped satellite gas will mix with hot host gas resulting in a high cooling rate (mixing layer cooling).  These processes can occur either within the satellites (less than 6 times gas scale radii) or in the wider CGM.
 

In these snapshots, we can see there is a lot of induced cool gas in the host CGM. Most of it, however, is spatially overlapping with the stripped cold gas and orange in color, indicating mixing. At later times, we can see in all of these diagrams that much of this cold gas (directly stripped from satellites and cooled from the halo) eventually goes within $40$ kpc and falls onto the host ISM.

The morphology of this cold gas in these different satellite distributions is quite different. We can see in these figures that the cold clouds from m10 are larger in size, whereas the less massive satellites of m09, m08 produce small cold clouds. The small clouds mix with the hot CGM and heat up in short time period as seen in the top right panel in Figure \ref{sch_m08}, where within roughly 1 Gyr all the clouds are destroyed and mix with CGM or have fallen into the central galaxy (as seen in the central top panel). However, even at 2 Gyr (right panel of Figure \ref{sch_m10}), the larger clouds of m10 survive, are clearly connected to the satellites, and continue to contribute to CGM cold gas budget.

The location of the satellite also plays an important role. Closer satellites not only feel more ram pressure due to the higher density of the CGM, but the cold clouds from them also fall faster on the host ISM. At earlier times, the left and middle panel of Figure \ref{sch_m10} do not show significant difference in cold gas around both the satellites. However, in the right panel of Figure \ref{sch_m10}, we can clearly see that the satellite at 150 kpc has more cold gas than the satellite at 100 kpc at later times $\sim2$ Gyr, as most of the gas from the closer satellite is stripped faster due to the higher ram pressure and the stripped gas falls within the $40$ kpc inner radius more quickly.  

Feedback also changes the morphology of the clouds. In the top panels of Figure \ref{sch_m09} and \ref{sch_m08}, we can see that the clouds are more dispersed (less dense) with more surface area than those in the bottom panels. These clouds have generally lost their coherent structure by the final panel (1 Gyr and 750 Myr in m09 and m08, respectively). However, in the bottom panels, it can be seen that the clouds are elongated and narrow without the energy from stellar feedback, hence offering less surface area and survive for a longer period of time. 
This difference between satellites with and without feedback is most clearly seen in the right panels of Figure \ref{sch_m08}: in the top panels all the cold clouds are dispersed and mixed with CGM by the 750 Myr, whereas in the bottom panel, there are still several cold clumps, with some of them falling inside $40$ kpc. We will discuss this in more detail later in Section \ref{S:feedback}.

\subsection{How much cold gas is there?} \label{S:amount}
In this section, we quantify the cold gas content of the CGM that is $\ge$ 40 kpc away from the center of the host galaxy. In Figure \ref{coldgasmass_tot}, we show the time evolution of total cold gas mass (See our definition of the cold CGM gas above in Section \ref{S:Define Cold Gas}) in the case of three different mass distributions of satellites: 1) 2 satellites of $2\times10^{10}\,M_{\odot}$ (m10: red), 2) 20 satellites of $2\times10^{9}\,M_{\odot}$ (m09: blue), 3) 200 satellites of $2\times10^{8}\,M_{\odot}$ (m08: green).  In Figures \ref{coldgasmass_tot}- \ref{pdf} the different linestyles denote different resolutions and the shading denotes distance from the host center.  Here we focus on the dark solid lines, and in later sections we discuss the other linestyles.

This total budget includes cold gas inside the satellite ISM, cold stripped gas from the satellites, and cold gas from the host CGM. The cold gas mass starts with nearly a value of $10^9 M_\odot$, as initially it only includes the satellite ISM. The total ISM mass of m10 is slightly different than the masses of m09 and m08 (see Table \ref{tab:ic}) in order to roughly keep the same baryonic fraction (as mentioned in  Section \ref{S:ic}) according to the stellar mass-halo mass relation of each halo mass.
Hence they start off with different values of the total cold gas mass. 
For the run with no satellites, there is very little cold gas mass (3 orders of magnitudes lower than the satellite runs) beyond $40$ kpc from the center of the host galaxy throughout the entire simulation. Therefore we conclude that in the runs with satellites, the dominant origin of cold  gas outside $40$ kpc is due to the cold gas inside the satellites, stripping from satellites and the induced cooling by the satellites.  
Over time, we can see a decline in the total cold gas budget in Figure \ref{coldgasmass_tot}. This is due to the fact that at late times, cold gas either falls into the host ISM or is heated by mixing with the CGM. We note that, the satellite ISM in m09 and m08 loses all of its gas at later times.

We study the cold gas in more detail by isolating all the different contributions in Figure \ref{coldgasmass1} where, we show the time evolution of instantaneous mass contributed by the satellites in the case of three different mass distributions of satellites. The top-to-bottom panels show the time evolution of mass of initial satellite cold ISM, stripped cold gas, host gas cooled inside of the satellite, and host gas cooled outside of the satellite respectively.  All the panels sum to Figure \ref{coldgasmass_tot}. In the first panel of Figure \ref{coldgasmass1},  we can see that the initial satellite cold ISM mass decreases with time as cold gas from the satellites get stripped via ram pressure. In the second panel of Figure \ref{coldgasmass1},  we show the time evolution of the stripped cold gas. At early times we see an increasing amount of cold gas being stripped from the satellites until the stripped cold gas mass reaches an early peak.  
Thereafter, the stripped cold gas mass starts declining slowly over time. There are two reasons for this decline. 
First, 
all cold stripped gas from the satellites that falls to within $40$ kpc is removed from our instantaneous CGM count. 
Second, most of the cold gas inside the satellites is stripped by this time, so additional cold gas is not being directly fed into the CGM. That is clear in the top panel of Figure \ref{coldgasmass1}, where the cold gas inside the satellite follows a declining trend with time as the cold gas gets stripped from the satellites and makes the satellite galaxies cold-gas deficient over time. 

In the third panel of Figure \ref{coldgasmass1}, we show the host gas that has cooled within the satellites. This is the hot host gas that falls inside the potential of the satellite and cools therein.  The time-scale of this mass evolution follows roughly the stripping time as once all the cold gas is stripped from the satellites, it is not possible for hot host gas to cool inside of the satellite.

Finally, the fourth panel of Figure \ref{coldgasmass1} shows the host gas that cools outside of the satellite in the mixing layer or because of turbulence-driven cooling.  Importantly, by comparing the second and fourth panel we find that the satellites induce a similar amount of cold gas as  the stripped cold gas from the satellites. Induced cool mass being proportional to (and not much larger than) the cold gas mass injected in the halo (e.g. by stripping) indicates that this induced cooling primarily happens by the mixing of cold stripped gas and hot CGM gas. This mixing brings the temperature of the host gas to a lower temperature where the cooling curve is higher, which can cause the hot gas to cool faster than the cold gas is mixed away. In our simulations this condensation of the hot CGM gas onto the cold cloud via the mixing layer does not runaway as is commonly seen in wind tunnel/cloud crushing simulations \citep{Gronke2018, Abru2022}. This difference is likely a result of the effect of gravity (from both the host and satellite) which leads to cold clouds infalling into the host ISM in our simulations (see for example \cite{Tan2023}).
In addition, the time evolution of the induced cold gas outside the satellite mimics the shape of the stripped cold gas with a slight delay time. This agrees well with our visual impression from Figures \ref{sch_m10}, \ref{sch_m09}, and \ref{sch_m08} that most of the induced cooled gas is located around the mixing layer.  

Together these findings imply that most of the induced cooling by the satellite occurs in the mixing layer of stripped cold gas from the satellite. Therefore, we argue that the prime satellite-induced mechanisms that can contribute to the cold gas budget of the CGM of host galaxies are ram pressure stripping and induced cooling in the mixing layer of this cold stripped gas.

\subsection{How do satellite properties affect the cold gas mass in the CGM?} \label{S:sat property}

In our suite of simulations we have varied several parameters:  satellite mass, stellar feedback, orbital distribution, number of satellites, and simulation resolution.  Here we discuss the impact of these variables on the cold gas content in the CGM of the Milky Way-like host.

\subsubsection{Dependence of cold gas mass on the mass distribution of satellites} \label{S:mass dist}

In the first panel of Figure \ref{coldgasmass1}, we see that the time evolution of cold satellite ISM gas follows a similar trend for different satellite masses. However, the rate at which the cold ISM gas disappears from the satellites is different for different mass distributions. While the cold ISM gas from m08 and m09 satellites disappears roughly within 0.25 Gyr and 1 Gyr, m10 satellites lose their cold gas at much slower rate, over more than a 3-4 Gyr timescale.  This is as expected given the higher restoring force in more massive satellites. 

A similar scenario also applies for the time evolution of cold stripped gas from the satellites (second panel of Figure \ref{coldgasmass1}). Although the general shape of the lines in the second panel of Figure \ref{coldgasmass1} for all three mass distributions is the same, the timescales of stripping are very different. The massive m10 satellites continue to feed cold gas to the CGM for a longer period of time ($\sim$ few Gyrs) than the less massive ones (m09: $\sim 0.5$Gyr, and m08: $\sim 1.5$Gyr). Therefore, we find that only massive satellites of at least SMC-like mass can contribute to the cold gas mass budget of the CGM for several Gyrs.  

In the third panel of Figure \ref{coldgasmass1}, we can see the maximum amount of host gas cooled inside the satellite is almost the same for m08 and m09 satellite distributions. From comparing the third and fourth panel, it is clear that more host-gas cooling happens outside of the satellites than inside of the satellites in the m09 and m08 runs. Only in the m10 runs is the amount of cold gas in the satellite a significant fraction of the total cold gas in the CGM, most likely due to the deeper potential well and larger net ISM mass for m10.



In the fourth panel of Figure \ref{coldgasmass1}, induced cooling of the host gas outside the satellites follows the stripped cold gas. As at early times, there is more stripping in m08 than m09 and m10 which leads to more induced cooling in the case of m08 initially. However, roughly after 0.25 Gyr, all the gas blows out of m08, whereas at that time, the stripping of cold gas peaks for m09. Thereafter, m09 induces more cooling outside of the satellite. The induced cool gas follows the stripped gas until the time the satellites become cold gas deficient. However, m10 continues to induce cold gas outside the satellite for several Gyrs even if at any given time the amount of cooled host gas is smaller than at the early times in m08 and m09.  This also reflects the slower stripping rate in m10.  

In summary, we find that a large number of low mass satellites can add a significant mass of cold gas to the CGM for a short period of time when they are initially stripped.  However, in order for cold gas from satellites to persist in the CGM there must be continuous feeding, which can only come from more massive satellites. 

\subsubsection{Dependence of cold gas survival time on the mass of satellites} 
\label{S:survival}
While we have shown that massive satellites (m10) continue to feed cold gas to the CGM of the host galaxy even beyond 4 Gyr while less massive satellites lose their cold gas quickly (within 0.5 Gyr and 1.5 Gyr  for m08 and m09 respectively), we have not yet attempted to carefully determine the fate of the stripped gas.  Here we examine whether cold stripped clouds remain cold for a long period of time or mix with the hot host CGM and get heated up. 

\begin{figure*}
\includegraphics[width=0.46\textwidth]{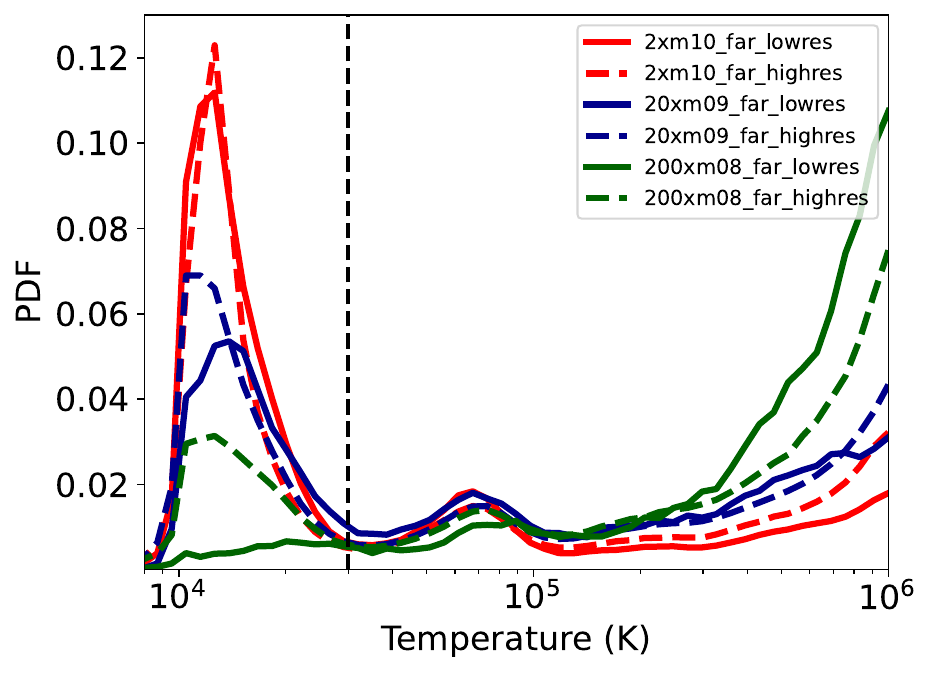}
\includegraphics[width=0.45\textwidth]{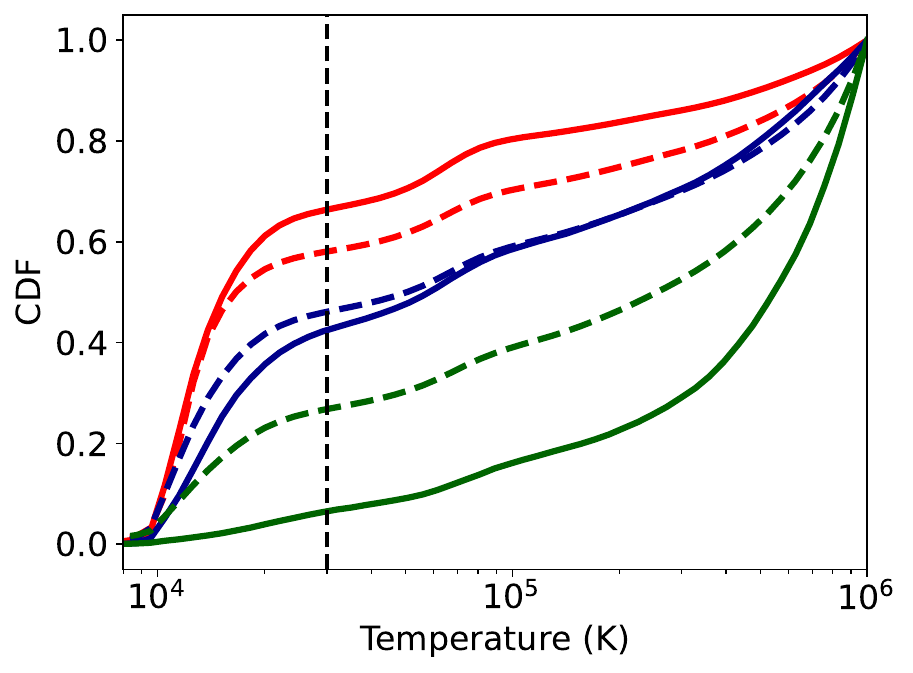}
\caption{The left and the right panels respectively show the time and mass-weighted probability distribution function (PDF) and cumulative distribution function (CDF) of cold-stripped cloud mass as a function of the temperature of the cloud after the stripping. The time over which the PDF and CDF are weighted is from the time of stripping to the time at which the cloud enters 40 kpc or the end of the simulation, whichever is earlier. }
\label{pdf}
\end{figure*}

To dig more deeply into these possible scenarios, we track the temperature evolution of all the gas that is stripped from satellites while cold. We begin tracking cold gas particles at the time of stripping, and continue until either the end of simulation or until the time the gas particle falls to within 40 kpc of the host center (which we define as leaving the CGM). We plot the time-weighted, mass-weighted probability distribution function of the temperature (PDF; left panel) and corresponding cumulative distribution function (CDF; right panel) of this stripped cold gas in the Figure \ref{pdf}.  

In this plot, we can see that the PDF for m10 has a large peak at a temperature of $10^4$ K along with one small peak at $10^6$ K. That means most of the gas that is stripped in the cold phase remains cold. The CDF quantifies that roughly 70\% of the cold stripped gas particles retain their temperature. For the PDF of m09, we can see two almost equal peaks at $10^4$ and $10^6$ K showing that $40-50$\% of the cold gas remains cold and the rest is mixed and heated. Continuing to the lowest mass satellites, the PDF of m08 has a bigger peak at $10^6$ K, indicating that most of the cold stripped gas is being heated before either the end of the simulation or before the cold clouds can fall to within 40 kpc of the host galaxy. We will discuss the impact of higher resolution (dashed lines) in Section \ref{S:resolution}.


Now we consider why we see such different trends of cloud destruction in different satellite distributions. For this, we estimate the ratio of cooling time (t$_{cool}$) to cloud crushing time (t$_{cc}$) using the satellite ISM as our ``gas clouds'', and accounting for the variation in galacto-centric distance by using the following cloud crushing equations \citep{Klein1994,Gronke2018,Gronke2020}:  
\begin{equation}
    t_{cool} = \frac{2.5 \times (k_b \times T_{mix})^2 \times ({\mu_{H}}/{\mu})^2 } {P_{\rm CGM} \times \Lambda(n_{mix},T_{mix},Z)}.
\label{tcool}    
\end{equation}
and 
\begin{equation}
    t_{cc} = \sqrt{\frac{T_{CGM}}{T_{sat}}} \times \frac{R_{scale} }{v_{sat}} = \sqrt{\frac{\rho_{sat}}{\rho_{CGM}}} \times \frac{R_{scale} }{v_{sat}}.
\label{tcc}    
\end{equation}
where temperature and density are inversely proportional at constant pressure.  Here R$_{\rm scale}$ is the gas scale radius of satellite, which is proportional to satellite mass, M$_{\rm sat}^{1/3}$, and $T_{mix} = \sqrt{T_{sat} \times T_{CGM}}$ \citep{Begelman1990}. Using the cooling rate from \cite{Coolcurve}, and using $T_{sat} = 10^4$ K, and $v_{sat}= 200$ km/sec (taken from the simulation values), we find that this ratio is much larger (a factor of $\sim 8$) in m08 than in m10. Unsurprisingly, using straightforward cloud crushing equations on the initial satellite properties we would expect the smaller m08 satellites to mix faster than the large m10 satellites. 

However, treating satellites as monolithic clouds is an oversimplification.  A visual inspection of Figure \ref{sch_m10}, \ref{sch_m09}, and \ref{sch_m08} clearly shows that the stripped gas from galaxies consists of smaller clouds with a range of sizes.  While the projections indicate that the massive, large m10 satellites produce bigger clouds whereas the clouds from less massive m09, m08 satellites are smaller in size, we can quantify this impression here.   

\begin{figure}
\includegraphics[width=0.48\textwidth]{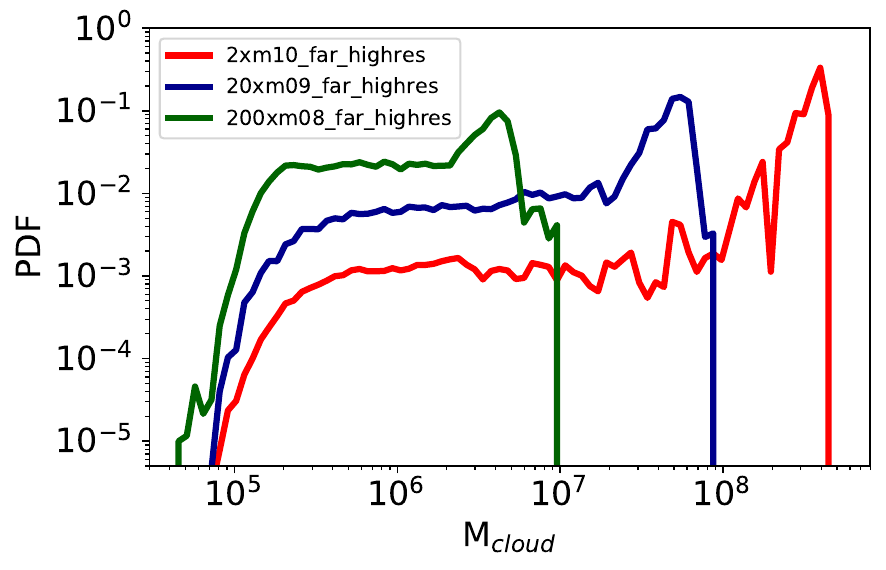}
\caption{{The time-weighted mass distribution of the cold stripped clouds for three different satellite distribution. This distribution includes the satellites, which is indicated by peak at higher mass of the distribution, while the lower-mass tail of the distribution is from the stripped cold gas from the satellites. While m09 and m10 satellites contribute massive clouds to the CGM, the clouds from m08 satellite is less massive.}}
\label{cloud_size}
\end{figure}

We smooth the gas particle data into  $256^3$ regular cells with an SPH-like deposition according to the particle smoothing length. We mass-weighted the temperature in each grid while depositing. We use skimage.measure\footnote{https://scikit-image.org} module (label, regionprops) of python to identify individual clouds in grid outputs of our simulations where the cloud-finder finds the continuous cells with temperature lower than $10^5$K. While saving data to the grid, it’s smoothing the temperature from the ``gas particles'' that overlap with the grid. Therefore, all grids are typically warmer than the original cold gas around the position. To reduce the statistical noise and avoid underestimating the cold gas, we make the temperature $10^5$K for the cloud finding calculation. In Figure \ref{cloud_size}, we show the time integrated mass PDF of these clouds. Note that we did not attempt to exclude the satellite itself from ``cloud'' identification, hence this PDF includes the satellite ISM, which is indicated by the peak at the side of higher mass of the PDF. Importantly, the PDF of each simulation shows a smooth distribution of cloud masses below the satellite mass.  
As predicted from the qualitative picture, the PDF for m10 satellites contains more massive cold clouds, whereas the cold clouds from m09, m08 satellites are less massive. The mass of the clouds from m08 satellites is roughly two orders of magnitude less than the mass of the clouds from m10. 
We verified that the cloud volume as measured on this grid is also largest in m10, and is related to cloud mass. Hence we can say that the mass/size of the clouds from a satellite is roughly proportional to mass/size of the satellites. 
The smaller, less massive clouds from m08 lead to smaller mixing times indicated by Equation \ref{tcc}. Hence these clouds are more easily heated up in a short period of time. On the other hand, bigger clouds from m10 have larger mixing time, therefore, surviving longer time by retaining their temperature. Hence, whether we use the satellites to determine cloud properties or measure the individual clouds stripping from the satellites, m08 is smaller than m10 and mixes faster. 

While there is much more physics included in our simulations, such as self gravity, star formation and feedback (See Section \ref{S:feedback}), it is reassuring that the size of the cold gas ``cloud" nevertheless correlates well with its survivability.  We find that cold gas stripped from low mass satellites (m08) forms smaller clouds that can be heated by the surrounding CGM, while cold gas stripped from high mass satellites resides in more massive clouds that persist for several Gyrs or until the clouds fall into the host galaxy.

\subsubsection{Dependence of cold gas mass on spatial location of the satellites in the CGM} \label{S:location}
In addition, we vary the distance of the satellites from the center of the host galaxy. In the Figure \ref{coldgasmass1}, the lighter shade color indicates the runs with satellite galaxies distributed closer to the center of the host. A satellite that is  closer to the center of the host feels more ram pressure than the farther one as the density of the CGM is inversely related to halocentric radius. Therefore, more gas should be stripped from galaxies closer to the halo center. This effect can be identified in Figure \ref{sch_m10} (rightmost; 2 Gyr snapshot), \ref{sch_m09} (top right; 1 Gyr snapshot), and \ref{sch_m08} (top middle; 250 Myr snapshot), where we can see that closer satellites lose their gas faster than the farther ones. However, another competitive fact which plays a role here is that the stripped gas from the closer satellites falls to within 40 kpc of the host galaxy faster. Therefore, as time progresses, the stripped gas from the closer satellite will fall faster within 40 kpc and therefore the additional stripped gas is not being accounted for in the instantaneous stripped cold gas measurements. 

We most clearly see these competing effects in the m10 runs.  In the top panel of Figure \ref{coldgasmass1}, we see that the cold gas inside the closer satellites is always less than in the more distant pair.  In the second panel we see that the stripped cold gas increases more quickly in the run with the closer satellites, due to the faster gas removal observed in the first panel.  However, because gas from the closest satellite (50 kpc from the host center) quickly falls to within 40 kpc from the host center, the amount of stripped cold gas in the CGM from the closer satellites flattens early, and more stripped cold gas is found in the CGM from the more distant satellites.  The mass of host gas that cools either inside or outside the satellites follows similar trends to the cold satellite gas mass.

These trends with satellite distance are independent of satellite mass, and in all cases we expect that cold gas clouds (of equal mass) that are stripped from more distant satellites will enhance the CGM cold gas budget for a longer time due to the longer infall time.


\subsubsection{Dependence of cold gas mass on stellar feedback} \label{S:feedback}
In this section, we investigate how stellar feedback from the host galaxy and satellites affects the contribution of cold gas to the CGM. In our run with no satellites, we find that stellar feedback from the host can contribute very little to the cold gas budget of the CGM, about 3 orders of magnitude less than what satellite galaxies would contribute. Additionally, we note that this contribution is not continuous, as cold gas in the CGM appears randomly for small duration ($\sim$100 Myr) due to the stellar feedback from the host. This is due to the fact that there are no large-scale winds from the host, consistent with FIRE galaxy formation simulation for MW-mass galaxies \citep{Pandya2021, Stern2021, Muratov2015}.

\begin{figure}
\includegraphics[width=8cm,height=21cm]{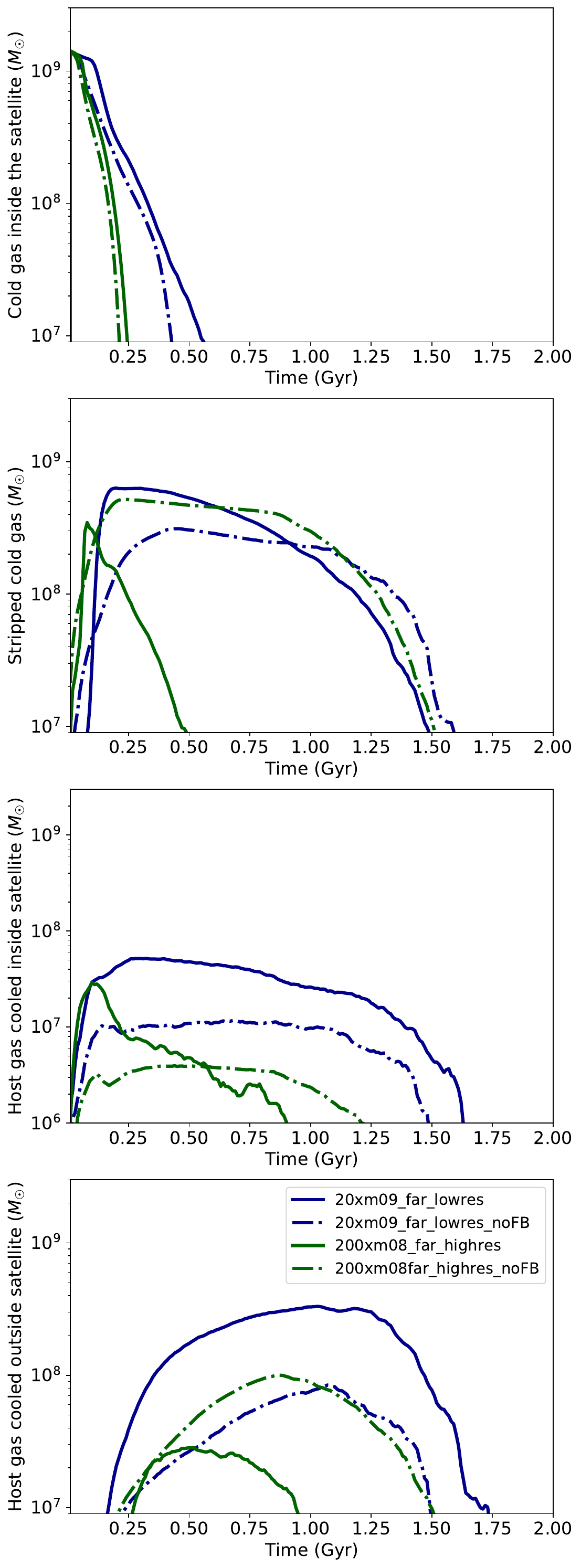}
\caption{{The time evolution of the cold (T$\le3\times10^4$K) gas mass from different contributions beyond $40$kpc radius from the center of the host galaxy for the case of no feedback in satellites and host along with no gas in the satellites for the $10^9$ and $10^8$ M$_{\odot}$ satellites. From left to right, the panels respectively describe cold gas stripped from the satellites which are cold satellite gas that falls beyond 6 times the scale radius of the satellites, cold gas induced inside of the satellites, and the host gas cooled outside of the satellites.}}
\label{coldgasmass_fb}
\end{figure}

However, the stellar feedback in satellite galaxies has a significant effect on the  gas added to the CGM of the host galaxy \citep{FG2016, AA2017} and the temperature of this gas. To study this effect, we investigate two of our runs with no feedback in the satellites and host, which are shown in Figure \ref{coldgasmass_fb}. From the top panel of Figure \ref{coldgasmass_fb}, we can see that there is little effect of feedback on the cold gas inside the satellite, which implies the rate at which the cold gas disappearing from the satellites is the same in both the feedback (fb) and no feedback (nfb) runs. 


However, feedback has various impacts on total gas stripping. Firstly, it amplifies the total gas (hot+cold) stripping process, resulting in a greater extent of gas removal. Secondly, it alters the size of the cloud, causing it to become more fragmented and smaller as time progresses. Lastly, feedback induces a rise in gas temperature. While the first effect leads to a higher mass of cold-stripped gas in the presence of feedback compared to the absence of feedback, the latter two effects counteract this by diminishing the mass of cold-stripped gas, particularly in later stages and for smaller satellites.

We can see in the second panel of Figure \ref{coldgasmass_fb}, for m09, at earlier times (before $\sim 1$ Gyr), there is more stripped cold gas from the satellites in the case of feedback than in the case of no feedback. This is due to the fact that the feedback removes satellite gas of all phases more efficiently. Along with cold gas, feedback also removes hot gas from the satellites, which can later cool down in the stripped gas tail and add to the cold gas budget to the CGM, hence reaching a higher peak in the feedback case. However, for m08, the potential well is so shallow that feedback blows out all of the satellite gas in a very short period of time ($\sim 0.2$ Gyr, see top middle and top right panels of Figure \ref{sch_m08}). Therefore, the stripped gas does not get enough time to cool down in the tail, and also, the above-mentioned effects two and three can dominate in small m08 satellites. Therefore, the early peak in stripped gas mass in the m08 feedback case is not higher than the peak stripped cold gas mass in the m08 no feedback case.

However, with feedback, the stripped cold gas in m09 and m08 survives for a shorter time than in the no feedback case. This is because feedback also changes the morphology of the stripped gas. Contrary to the feedback case, cold clouds in the no feedback case are elongated and denser  (see second panel of Figure \ref{sch_m09}, \ref{sch_m08}). Therefore, without feedback, these denser and elongated clouds (Bottom panels Figure \ref{sch_m09},\ref{sch_m08}), will survive longer than the less dense, small clouds generated in the feedback case (following the Equation \ref{tcc}). This trend is stronger in the case of m08 than m09. 

\begin{figure*}
\includegraphics[width=0.49\textwidth]{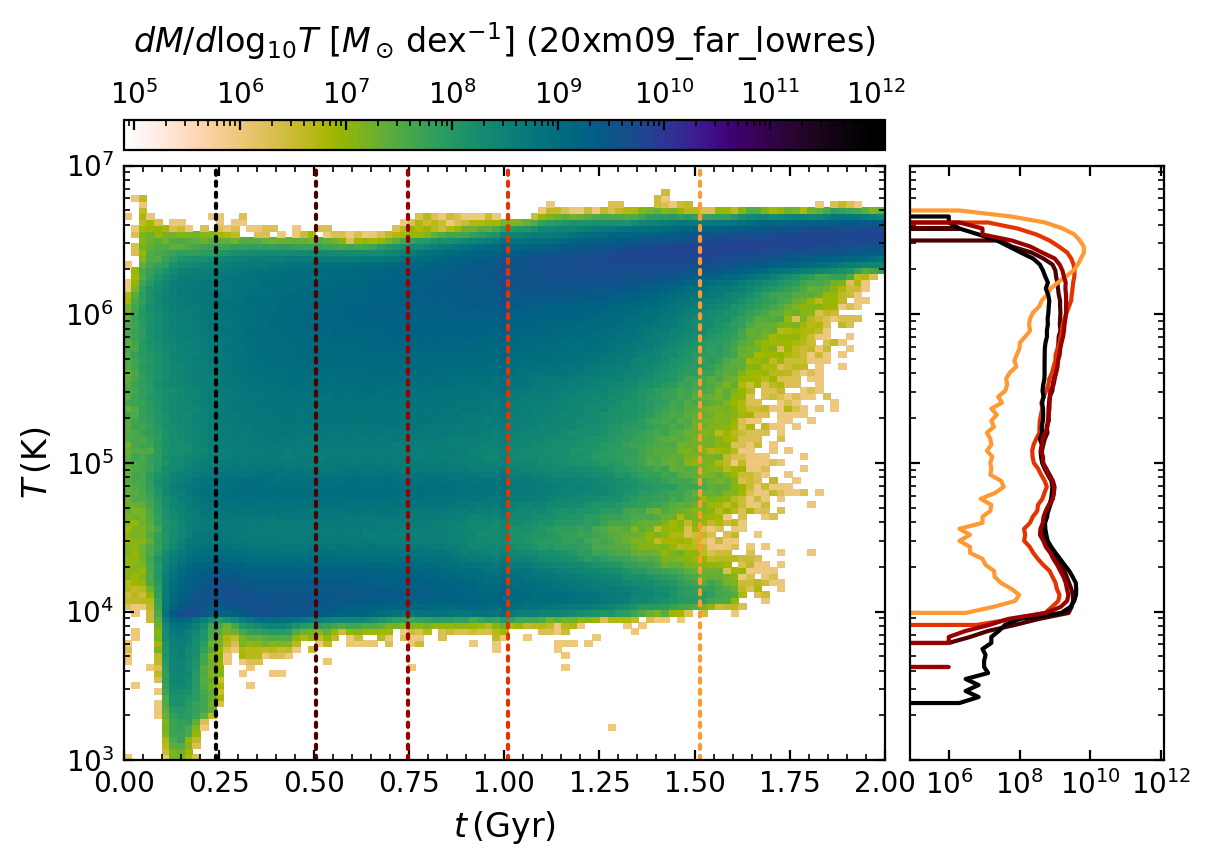}
\includegraphics[width=0.49\textwidth]{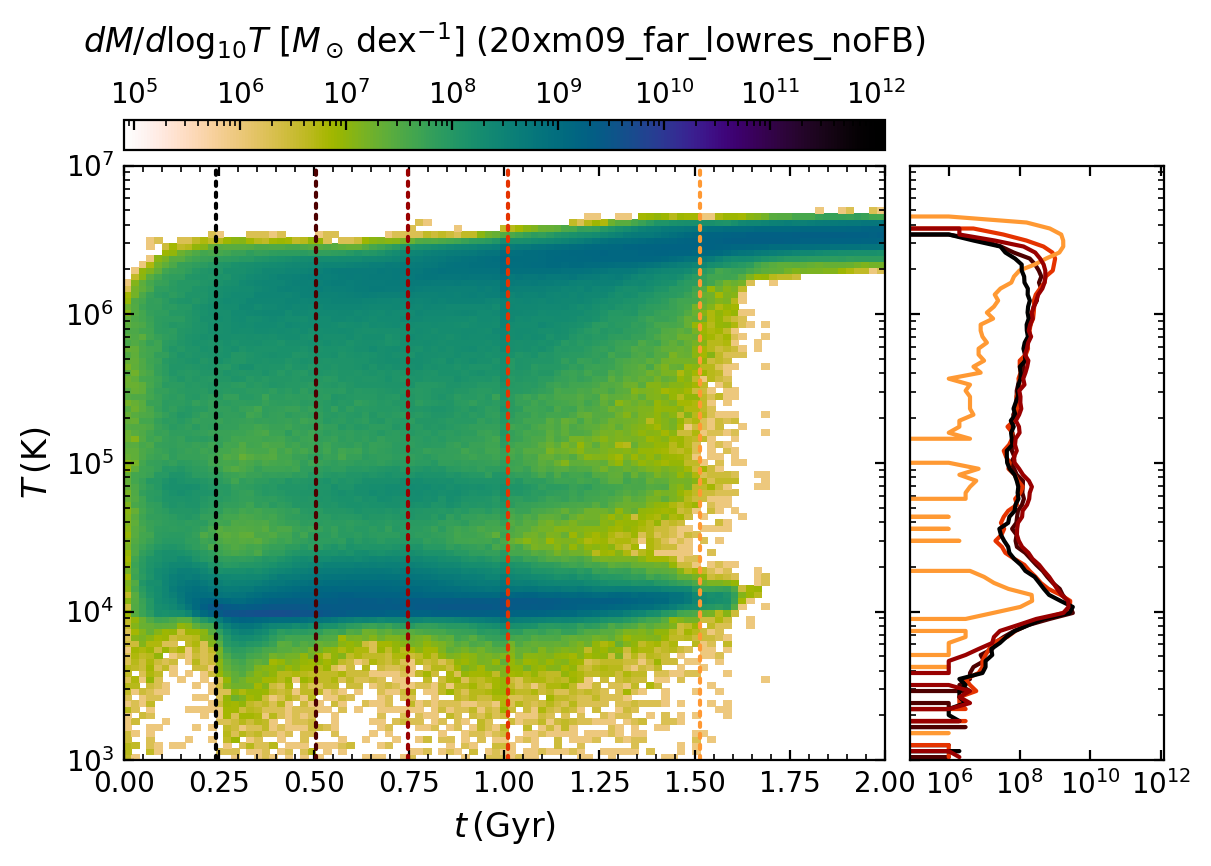}

\includegraphics[width=0.49\textwidth]{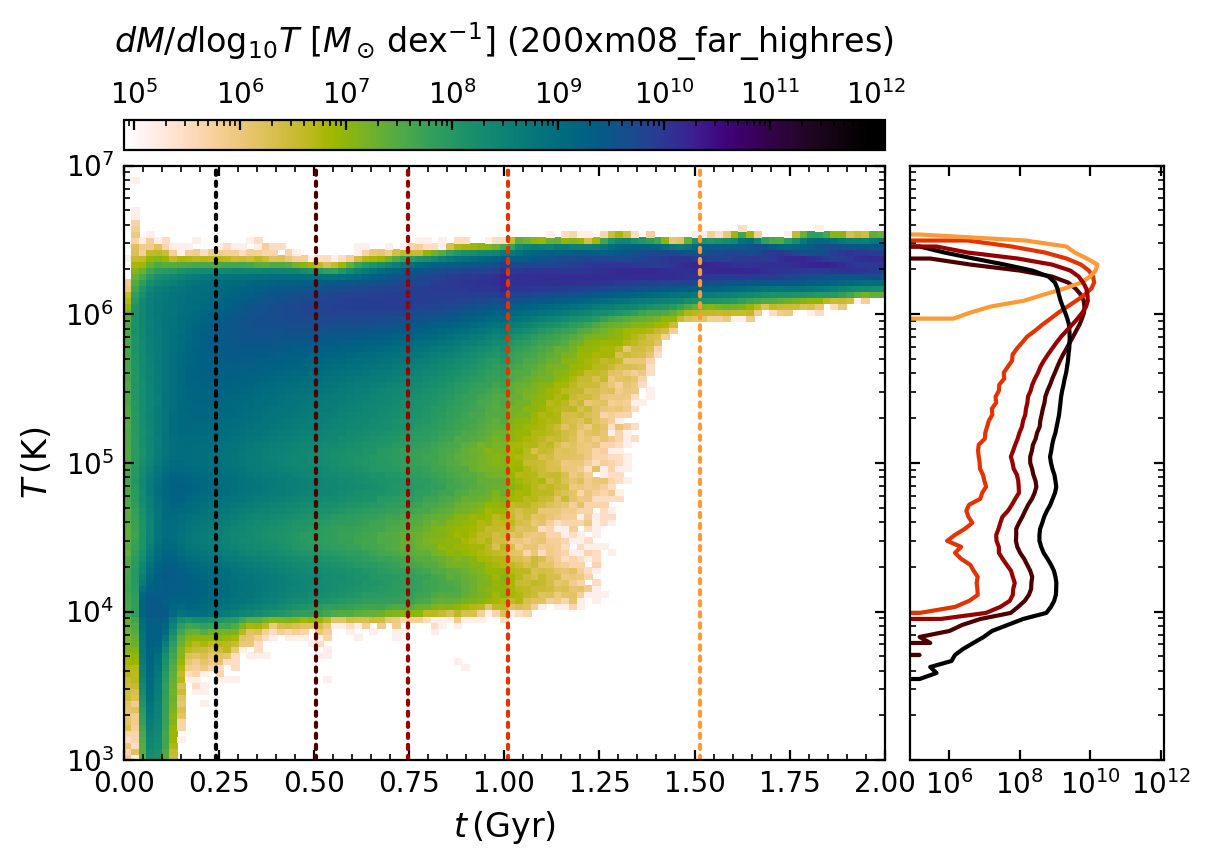}
\includegraphics[width=0.49\textwidth]{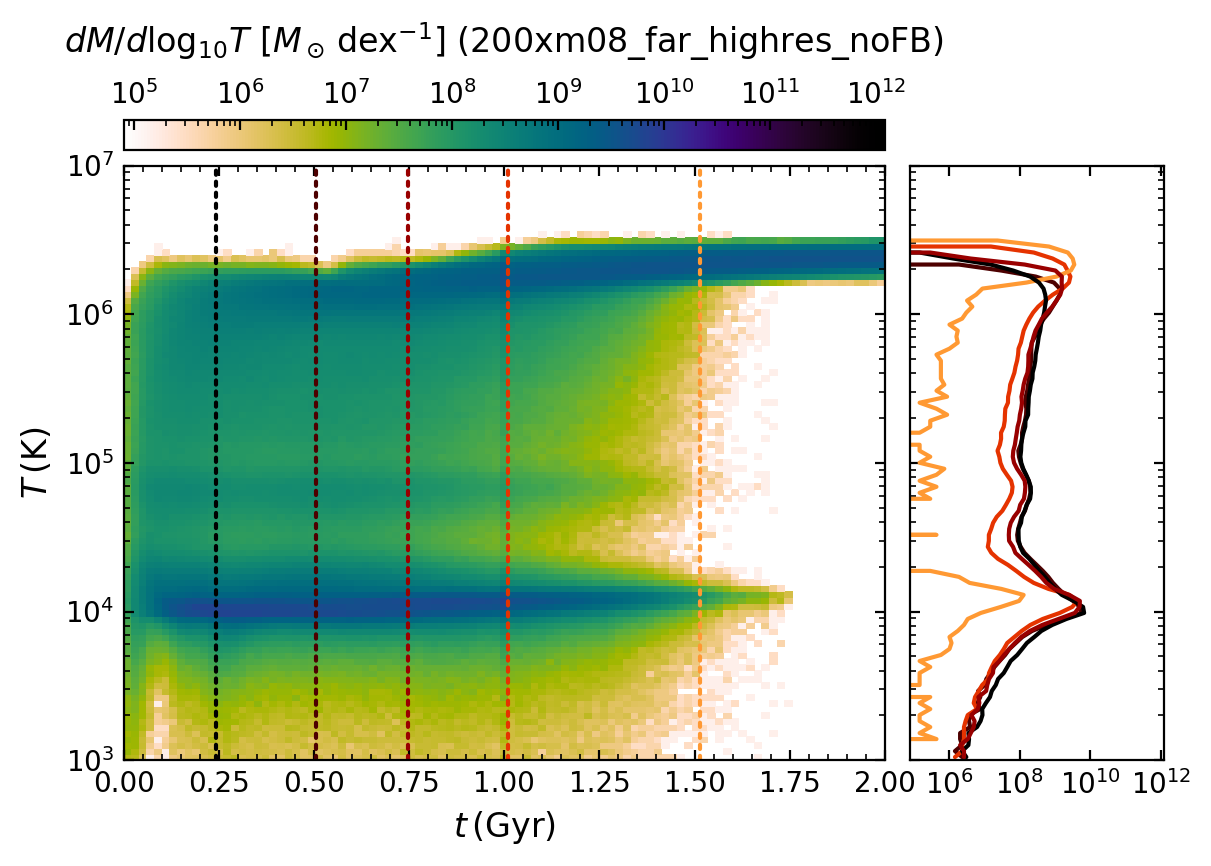}
\caption{The left and the right panels show 2-d probability distribution function (PDF) of stripped gas mass as a function of the time and temperature of the gas in the case of with feedback and without feedback  respectively for the m09 (top panel) and m08 (bottom panel) satellites. In the right bar of each 2-d histogram plot, we show the 1-d histograms of temperature at single snapshots denoted by the similar colored vertical lines in 2-d histograms.}
\label{2d-pdf}
\end{figure*}

This is also evident from Figure \ref{2d-pdf}, where we show the time evolution of the mass weighted temperature probability distribution (PDF) of the stripped gas from the satellite as a function of time and temperature of the gas. The top and bottom panels respectively indicate the m09 and m08 runs whereas left and right panels denote the cases with and without feedback, respectively. It is clear from the plots that, with no feedback the horizontal strip of cold gas distribution at $10^4$K exists for a long time, until $\sim1.6$ Gyr for both m08 and m09, whereas, with feedback, gas gets more sparse and smoothly distributed across all temperatures. The difference is the most dramatic in m08, with the cold stripped gas in the feedback run being destroyed in a very short time of $\sim1$Gyr. In the right bar of each 2-d histogram plot, we show the 1-d histograms of temperature at single snapshots denoted by the similar colored vertical lines in 2-d histograms. We can infer the same scenario from these 1-d PDFs. For example, in m08 case with feedback, at 1.5 Gyr in the 2-d histogram (orange vertical line), there is no cold gas. However, without any feedback, m08 case shows cold gas at the same snapshot. 
Moreover, until 1 Gyr, all the 1-d pdfs of temperature in the case of both m08 and m09 look similar in no feedback case, indicating that most stripped gas does not change temperature. On the other hand, with feedback the low temperature distribution of 1-d PDFs shifts to higher temperatures, indicating mixing-driven heating of the cold stripped gas.   


From the third panel of Figure \ref{coldgasmass_fb}, we can see that the feedback induces more host gas cooling inside both the m08 and m09 cases than no feedback cases, however, for m08, feedback soon blows out all the gas from the satellite system retaining no gas to induce host gas cooling. Therefore, induced cooling of host gas inside the satellites end faster in m08 feedback case than the no feedback case.

As mentioned earlier, host gas outside of the satellite gets cooled in the mixing layer of stripped cold gas, hence, it roughly follows the stripped cold gas. For m09 with feedback case, there is more stripped gas than no feedback, hence, there is more mixing layer cooling (See the fourth panel of Figure \ref{coldgasmass_fb}). However, for m08, the trend is opposite as in m08 feedback case,  there is no stripped cold gas in the CGM within a very short period of time ($\sim 0.5$ Gyr) due to the rapid stripping of the gas by feedback and low cloud destruction time. Hence, there is no cold gas retained in the stripped tail to induce cooling in the mixing layer. However, with no feedback, since there is no rapid blowing out by feedback and the clouds are longer and denser, they tend to survive for a longer time (see Figure \ref{sch_m08}). We also note that the clumps in the no feedback run drop out of the cold gas budget by falling within the central $40$ kpc rather than being heated. 
 We have verified that cold gas stays cold in the m08 and m09 no feedback runs using the same analysis as in Figure 6 (not shown). 

To summarize, feedback not only controls the temperature of the gas added to the CGM, but it also controls the morphology of the stripped clouds. Along with cold gas, feedback removes hot gas from the satellite which can cool down and add to the cold gas budget of the CGM. Without energy from feedback, the clouds are elongated and denser than the feedback case, for which they maintain their coherent shape for a long period time and survive longer than the dispersed less dense clouds in the feedback cases.

\subsubsection{Dependence of cold gas mass on the number of satellites} \label{S:number}
\begin{figure*}
\includegraphics[width=1.0\textwidth]{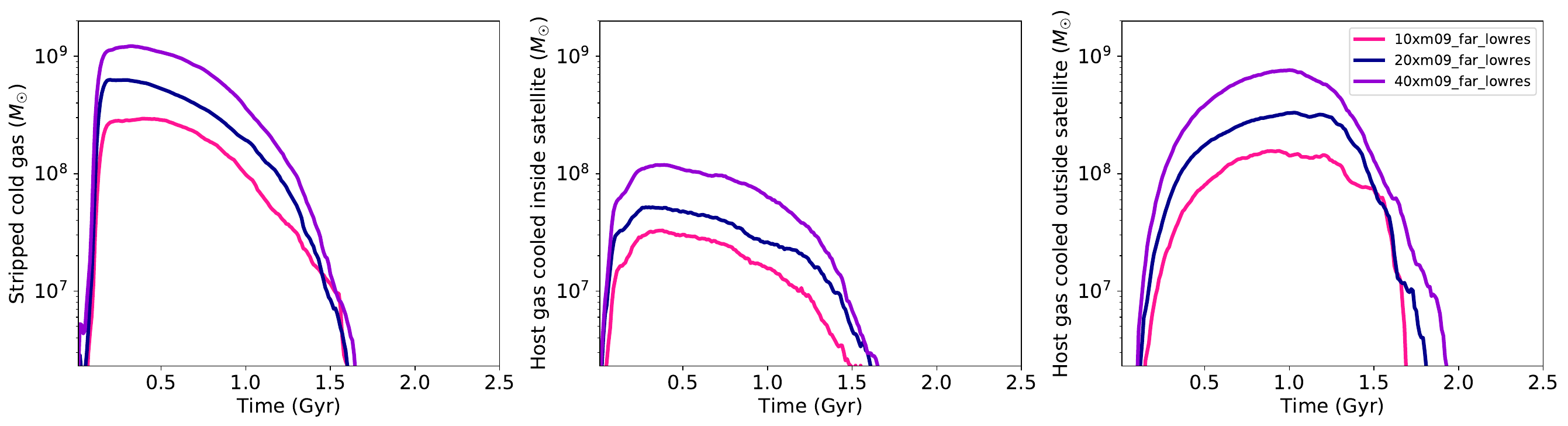}
\caption{{The time evolution of the cold ( T$\le3\times10^4$K) gas mass from different contributions beyond $40$kpc radius from the center of the host galaxy for the change in a number of satellites in the case of $10^9$ M$_{\odot}$ satellites. From left to right, the panels respectively describe cold gas stripped from the satellites that are cold satellite gas which falls beyond 6 times the scale radius of the satellites, cold gas induced inside of the satellites, and the host gas cooled outside of the satellites.}}
\label{coldgasmass_num}
\end{figure*}

We also investigate how changing the number of satellites affects the stripped gas or induced cooling contribution to the CGM. In Figure \ref{coldgasmass_num}, we show three cases of m09 run with 10, 20, and 40 satellites in pink, blue, and purple, respectively. The total amount of gas mass in the satellites in the system directly corresponds to the number of satellites. As we would expect, the contribution of stripped cold gas increases with the number of satellites (first panel) due to increase of gas mass in the system. We can see from the plots, the 40 satellites run has 2 times and 4 times more stripped cold gas than the 20 satellites and 10 satellites respectively, which is roughly linear in relation. 

Given that we expect the cooling of the hot CGM gas to be dominated by mixing layer cooling, we also expect that the host gas cooled, either inside or outside the satellite, should also be related to the number of satellites. Indeed, this is what we find in the second and third panels. The increase in cold host gas is directly related to the increase in stripped gas. Therefore, increasing the number of satellites does not enhance the cold gas mass of the CGM beyond the direct correlation with the total gas mass in the satellites.  

\subsubsection{Dependence of cold gas mass on the resolution of simulations} 
\label{S:resolution}
We have also performed higher resolution runs, which are shown by dashed lines in Figure \ref{coldgasmass1}.  We expect that the high-resolution (hr) runs will better resolve more dense, cold gas than low resolution (lr) runs, and this can result in either harder to strip gas or more cooling in the stripped tail, which will have competing effects. 
 Upon immediate inspection, we find that the cold gas mass in the hr runs is qualitatively similar to the lr runs and shows the same trends with satellite mass, indicating that our general results are robust to resolution. 

However, running with higher resolution does not affect every simulation in the same way.  For example, in the m10 case, initially stripping cold dense gas in the hr run is more difficult, therefore hr shows less cold stripped gas in the beginning than the lr run (second panel). Whereas, at later times, both resolutions show a similar amount of cold gas. This is due to the fact that at late times, when there is a significant amount of stripping, the stripped hot gas can radiatively cool more and the stripped cold gas can remain cool in the hr run due to its higher density than in the lr run. We would also expect there to be more cold gas inside the satellite for hr, as less gas is stripped initially and more satellite gas can radiatively cool inside the satellite due to the higher density in hr run.
This effect is more strongly seen in the m10 run, and seen to a lesser degree in the m09 run. 


{We briefly note that the bottom-most panel of Figure \ref{coldgasmass1} shows an increase in the `hot gas cooled outside satellite' for the higher resolution run of the m09 simulation at later times, around t=3 Gyr, that is not seen in the standard resolution run. On close inspection, we find that a very small number of particles (total mass of $\sim 3.2\times10^5\, \rm M_{\odot}$) show similar behavior in the low-resolution run, which may be because dense structures are not as well resolved as high-resolution run(as we have mentioned above). Tracking these gas particles, we find that they get denser and cool down rapidly within 100 Myr when they fall close to 40 kpc. We consider it to be more likely that this is due to interactions with the host stellar feedback rather than with stripped material from the satellites. However, as most of this cooling happens right around our radial cut, which is an arbitrary cut, we do not want to read too much into this later peak of cold gas.} 

For the least massive satellites of m08, in both lr and hr cases, the gas stripping occurs very rapidly and a lot more quickly than from the m10 and m09 satellites due to their weaker gravitational potential. 
 The difference between hr and lr runs for the cold gas inside the satellite is not significant (for a significance study see the Appendix \ref{appendix}) as they have very short stripping times and do not have enough time to cool more gas inside the satellite. However, later on, more satellite gas is able to cool in the stripped tail of hr runs for m08.

Survival of cold stripped gas in the hr runs also follows a similar story as in the lr runs (See Figure \ref{pdf}). However, one can clearly see a higher $10^4$K peak in hr runs than lr for m08 and m09. This implies that the cold gas in hr runs, which better resolves more dense gas, is retaining its cold temperature more than in the lr runs.

In addition, for all the satellite distributions, the host gas that cooled inside the satellite does not show a significant difference between the hr and lr cases. Importantly, the host gas cooled outside of the satellite follows the trend of the stripped gas mass in both lr and hr cases for all the satellite distributions. Our result holds that the more stripped gas there is, the more gas will be cooled in the mixing layer of the stripped gas.

\section{Discussion}
\label{S:discussion}
\subsection{How much condensation is induced by the satellite driven turbulence?} \label{S:turb}
We have discussed in the earlier sections that host gas cooling can be induced by the satellites in two ways: by mixing layer cooling and by turbulence-driven cooling. 
 We have argued that mixing-layer cooling better matches our distribution of induced cold gas, but here we examine this question in more detail.  
 
 First of all, we can see in Figures \ref{sch_m10}, \ref{sch_m09}, and \ref{sch_m08} that most of the induced cold host gas is spatially around the mixing layer of stripped satellite gas. Hence, we have run one case with no gas in the satellite to distinguish between the contribution of induced cool gas by these two processes. In this case there is no gas to be stripped and consequently no cooling in the mixing layer of this stripped gas. Therefore, if there is any induced cold gas outside the satellites it must be contributed by turbulence-driven cooling. We do not find any induced cool gas outside the satellite in this no satellite gas run. 
 
\begin{figure}
\includegraphics[width=0.5\textwidth]{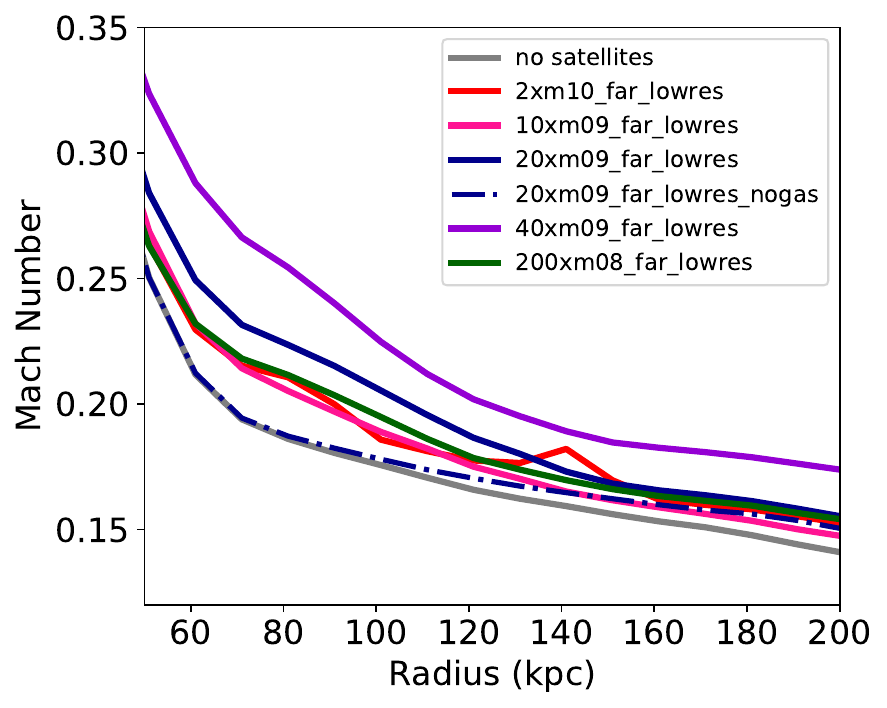}
\caption{{This plot shows the radial profile of time-integrated Mach number. It is significantly smaller than unity in all runs, implying subsonic velocity dispersion in the CGM of host galaxy even with the inclusion of different satellite distributions.}}
\label{turb}
\end{figure}

In addition, we calculate the turbulent Mach number. {We have calculated Mach number by taking the square-root of the sum of the variance of each velocity component and dividing it by velocity of sound. Taking the standard deviation of each velocity component has removed the average velocity, by excluding radial inflow, rotation, etc.} We show the radial profile of time integrated Mach number (over the timescale of 0.5 Gyr to 1.5 Gyr) for different satellite distributions in Figure \ref{turb}. The Mach number for different satellite runs are very similar to the runs with no satellites and with no gas in the satellites. Moreover, Mach number in each case has small range of values and is much less than one, implying the velocity dispersion in the host CGM remains always subsonic. Subsonic turbulence induces small density perturbations, which do not cause cold gas to precipitate out of the hot phase \citep{Balbus1989, Stern2019, Esmerian2021}. Therefore, we conclude that the induced cooling outside the satellites mainly happens in the mixing layer of the stripped cold gas from the satellites and there is not much contribution of turbulence-driven cooling. 


\subsection{What do the observations of Milky Way tell us?} \label{S:obs}
Our idealized simulation does not incorporate the realistic satellite distribution of the Milky Way galaxy. However, the mass range of satellites that we cover spans the high-end mass range of Milky Way satellites. We now consider how our estimation of cold gas from the different satellites compares with the observed cold gas budget of the Milky Way CGM. 

In their study, \cite{Putman2012} made an estimation of the overall mass of cold gas in the Milky Way halo that is identified through High-Velocity Clouds (HVCs). They determined this mass to be approximately $2.6\times10^7$ M$_\odot$, excluding the Magellanic Stream system. They also account for the presence of ionized components and helium and the estimated mass of the cold gas should be approximately doubled and multiplied by 1.4, resulting in a total mass of approximately $7.4\times 10^7\, M_{\odot}$. However, when including the contribution from the Magellanic Stream, which is not necessarily a common feature of all galaxies, the total mass increases by around $3\times 10^8\, M_{\odot}$. 

In another recent work by \cite{Rich2017}, the authors utilized data on ion covering fractions, previously determined distances and metallicities of HVCs, along with measurements of total Silicon and Carbon column densities in HVCs. Based on this information, they derived an estimate for the combined gas mass of the neutral and ionized CGM of the Milky Way, as traced by HVCs, which amounts to at least $3.0 \times 10^9$ M$_\odot$.  

Now let us see whether our estimated total cold gas brought by different satellites is within the range of the observational findings. 
The observed cold gas mass estimates by \cite{Putman2012} roughly agrees with the cold gas mass contributed by an m10 satellite (about SMC/LMC mass, \cite{Besla2015}): $\sim10^8\, M_{\odot}$ for stripped gas and $\sim2\times10^8\, M_{\odot}$ for total induced gas (see second to fourth panels of Figure \ref{coldgasmass1}). On the other hand, the contribution by a single m09 satellite matches this observation only by $\sim10\%$: $3\times10^7\, M_{\odot}$ for stripped gas and $\sim10^7\, M_{\odot}$ for total induced gas. However, if we compare to \cite{Rich2017}'s result, which has a higher estimate, we find that our calculated cold gas mass from m10, m09 satellite matches only 10\% and 1\% of the observational lower limit respectively. The contribution from m08 however is insignificant as m08 immediately ($0.25$ Gyr) blows out all of its gas.

We also highlight that the timeline for contributing cold gas is $\sim1$ Gyr for m09, whereas m10 can distribute cold gas even after 3.5 Gyr. This brings the question what is the infall times for the satellites of MW. From recent studies by \cite{Rocha2012,Fill2019}, the infall times of MW satellites vary from 1 Gyr to even 10 Gyr, as for example, the LMC/SMC has fallen into the MW potential roughly 1.5-2 Gyr ago \citep{Patel2017}. This implies that the LMC and SMC can be important sources of cold gas for the Milky Way CGM. They can not only provide cold gas to the Milky Way CGM from the time of its infall, but also bring in the observed budget of cold gas to the Milky Way CGM. 

Note that ten m09 satellites can also bring in a similar amount of cold gas to the Milky Way CGM, however, their contribution to the cold gas budget of MW CGM is likely to be short-lived (at or below 1 Gyr). Hence, if roughly ten m09 like satellites have fallen to the MW within past 1 Gyr, they can definitely contribute the observed amount of cold gas to the MW CGM. While infall times of different satellites vary, \cite{Lovell2023} (See their table 1) list fifteen satellites at or above the m09 mass in the MW (circular velocity greater than 16.4 km/s), indicating that satellites could have added a major fraction of the cold gas of the halo.
 
However, our calculation only gives a rough estimate of cold gas budget in the CGM for the following reasons: 1) it does not take into account realistic satellite distribution of Milky Way, 2) it does not consider the CGM of the satellites which can lead to more ram pressure stripping, and 3) it does not take into account other important components that can affect CGM cooling like cosmic filaments and AGN feedback.    

\subsection{What are the other components that can affect cooling of the CGM?} \label{S:other cooling}
Now one can ask a valid question: what are the other additional sources of cold gas in the outer CGM beyond satellite galaxies? 
Although this remains an open question, one possible mechanism is accretion through cold streams collimated by large-scale structure filaments.
In contrast to the hot virialized accretion mode, the unshocked cold $\sim10^4$ K gas can be transported through cosmic filaments into the galactic halos. Recent observations support this theory of cold mode accretion \citep{Dij2009, Goerdt2010, Ros2012, Daddi2022}. Galaxies close to cosmic web filaments experience an enhancement in star formation processes, as supported by studies such as \cite{Darvish2014,Vulcani2019}. Moreover, the studies conducted by \cite{Kotecha2022,Zheng2022} have provided evidence suggesting that filaments not only enhance star formation activity but also potentially delay quenching in galaxies. This enhancement in star formation could result from the accretion of cold gas fuel for star formation from cosmic web filaments.
Simulations also pose a similar story. A recent study by \cite{Hasan2023} pointed out that in high-mass central galaxies, there is a notable decrease in the gas fraction ($f_{\rm gas}$) at a distance of approximately 0.7 Mpc from the node (maxima of the density field), followed by a sharp increase at shorter distances. Although this work does not analyse only cold gas, this increase in total gas fraction points towards the accretion of gas from cosmic filaments. 

Another component which may play a prime role in the cooling of the CGM is Supermassive black hole (SMBH) or Active Galactic Nuclie (AGN) feedback. Feedback from SMBH/AGN has been suggested to have various impacts on the cooling of the CGM. AGN feedback, especially in the form of kinetic mode feedback at low rates of black hole accretion, has the effect of ejecting and heating up gas within and around galaxies. It acts as both an ``ejective'' feedback by expelling cold gas, as well as a ``preventative" feedback by increasing the average entropy and cooling time of the CGM \citep{Zinger2020, Somerville2015, Tumlinson2017}. In ``ejective'' feedback, AGN can physically expel cold gas from galactic disk. Some of this material may later fall back into the galaxy, contributing to the recycling of the CGM and enriching it with metals from the galactic center. This method can feed the CGM with cold gas from the disk. On the other hand, 
``preventive'' feedback from AGNs can inject energy into the surrounding material, causing temperature increases and resulting in the ionization of metals through collisions and photoionization \citep{Mathew2017,Mc2018,Oppen2018,San2019}. Multiple observations also suggest that the thermodynamics of the CGM are highly influenced by the energy released by AGN \citep{Nulsen2009, Werner2019}. When the CGM gas is heated through such mechanisms, the cooling of the CGM gas can be suppressed. In conclusion, AGN could both increase the cool gas mass of the CGM by ejecting material as well as suppress cooling of the CGM by heating it up.

\section{Conclusions and future plans} \label{S:conclusion}
We investigate the origin of the cold gas in the outer CGM and how satellites can impact the CGM cold gas budget over time. For this study, we have performed controlled experiments with a host galaxy of Milky Way mass along with satellite galaxies with three different satellite mass distributions (m10, m09, m08). Below, we list our main findings from this investigation.

\begin{itemize}
\item Satellite galaxies can contribute to the cold gas budget of the CGM of a MW-type host galaxy (Figures \ref{sch_m10}, \ref{sch_m09}, \ref{sch_m08}, and \ref{coldgasmass_tot}). The setup with no satellite galaxies produces three orders of magnitude less cold gas in the CGM than the runs with satellites. 

\item There are three main mechanisms by which satellites can add cold gas to the CGM in our simulations. The cold gas can be stripped from the satellite via ram pressure. Along with this, gas can also be removed from the satellites by feedback. Satellites can also induce cooling in the mixing layer of this stripped cold gas. We identify two mechanisms (direct ram pressure stripping and mixing layer cooling) that contribute similarly to the cold gas budget of the CGM
(Figure \ref{coldgasmass1}).

\item The spatial location of satellites also has a significant effect on stripping. Satellites closer to the host galaxy feel more ram pressure and are stripped faster due to the higher CGM density (Figure \ref{sch_m10}). For this reason, we see more stripped gas initially in the closer distributed satellites than the further ones. However, another competing effect is faster falling of stripped gas inside $40$kpc for closer-distributed satellites, which makes the stripped gas from them flatten earlier than the farther-distributed satellites (Figure \ref{coldgasmass1}). 

\item The contribution of cold gas by different satellite distributions are dramatically different, even when the total gas mass brought in by the satellites is the same. The less massive satellites (m08, m09) get stripped faster and lose all of their cold gas in a short period of time, while the massive SMC-like satellite (m10) continues to provide cold gas to the host CGM for several Gyrs. Therefore, only LMC or SMC-like satellites can add a cold phase gas mass of order $10^8$M$_\odot$ to the total cold gas budget of the MW-type host CGM for at least 4 Gyr (Figure \ref{sch_m10}, \ref{sch_m09}, \ref{sch_m08}, \ref{coldgasmass_tot}, and \ref{coldgasmass1}).

\item Different satellite distributions produce cold clumps of different size and mass. The less massive satellites produce smaller clouds with small cloud-crushing times that can easily be destroyed and heated. However, massive SMC-like satellites produce bigger clouds, which survive for a longer period of time (Figure \ref{pdf} and \ref{cloud_size}).

\item Stellar feedback from the host galaxy produces three orders of magnitude less cold gas than the satellite contributions due to the absence of large-scale winds from the host, in FIRE galaxy formation simulation for MW-mass galaxies. Furthermore, this contribution due to feedback from the host is not continuous; cold gas in the CGM appears randomly for brief ($100$ Myr) time periods and only at small radii. 
However, supernova feedback from satellites has a significant effect on the morphology of the cold gas. Feedback makes cold clumps more diffuse and increases their surface area, which not only induces more mixing layer cooling but also speeds their destruction. However, without feedback, cold clouds have smaller surface area producing less mixing layer cooling. These clouds are denser and survive longer (Figure \ref{coldgasmass_fb} and \ref{2d-pdf}).

\item An increased number of satellites linearly increases the stripped gas mass as well as the induced cool gas in the mixing layer. The total gas mass in the satellites is directly proportional to the number of satellites present in the system. Hence, the ram-pressure stripping along with induced cooling in the mixing layer are enhanced with this increased amount of cold satellite gas directly related to the increase in satellite number. (Figure \ref{coldgasmass_num}). 
\end{itemize}

In future work, we plan to incorporate a realistic distribution of the satellites of Milky Way with realistic orbits \citep{Santis2023} and to investigate the effect on the cold gas budget of the Milky Way CGM. The presence or lack of a satellite CGM in the initial conditions may also make a major difference \citep{LMC2022}. It is expected that isolated galaxies of similar mass as our satellites have more mass in their CGM than in their ISM \citep{Hafen2019}, and in an analysis of the fate of satellite CGM it is also found that much of it accretes onto the central galaxy \citep{AA2017, Hafen2020}. It is also seen that the satellite CGM plays a major role in the cooling of CGM gas of the more massive host in cosmological FIRE simulations \citep{Esmerian2021}. Other idealized simulations of satellite stripping that do not include the satellite CGM are unsuccessful in producing sufficient cold gas mass to match observations, and underestimate the importance of satellite galaxies for galaxy growth as a whole \citep{Bustard2018}. Our future plan is to include the CGM in the satellites and investigate the change in the amount of the cold gas contributed by the satellites. However, one can take our current estimates as a lower limit of the cold gas produced in the host CGM by the satellite galaxies.

In conclusion, when satellites bring in their own ISM to the CGM of the host galaxy, their ISM not only gets stripped but they also induce cooling in the host CGM. 
We universally find that at any given time, satellites induce about the similar amount of cold gas in the CGM as their own stripped ISM at that time. Therefore, satellites have a larger and dynamic impact on the cold gas in the CGM than a simple accounting of their cold gas mass would indicate.

%

\acknowledgments
We thank Greg L. Bryan, Rachel S. Somerville, Christopher C. Hayward for useful discussions and suggestions. We thank FIRE collaboration for useful discussions and suggestions. {We also thank the anonymous referee for the detailed comments which have been helpful to improve the manuscript.} MR acknowledges support from the Center for Computational Astrophysics (CCA) Pre-doctoral Program, during which this work was initiated. 
KS acknowledges support from the Black Hole Initiative at Harvard University, which is funded by grants from the John Templeton Foundation and the Gordon and Betty Moore Foundation, and acknowledges ACCESS allocations TG-PHY220027 and  TG-PHY220047 and Frontera allocation AST22010.
The computations in this work were run at facilities supported by the Scientific Computing Core at the Flatiron Institute, a division of the Simons Foundation.
CAFG was supported by NSF through grants AST-2108230  and CAREER award AST-1652522; by NASA through grants 17-ATP17-0067 and 21-ATP21-0036; by STScI through grant HST-GO-16730.016-A; and by CXO through grant TM2-23005X.

\vspace{0.3cm}
\section*{Data Availability statement}
The data supporting the plots within this article are available on reasonable request to the corresponding author. A public version of the GIZMO code is available at \href{http://www.tapir.caltech.edu/~phopkins/Site/GIZMO.html}{\textit{http://www.tapir.caltech.edu/$\sim$phopkins/Site/GIZMO.html}}.

\bibliographystyle{mnras}
\bibliography{mybibs}

\appendix

\section{significance study} \label{appendix}
\begin{figure}
\includegraphics[width=0.5\textwidth]{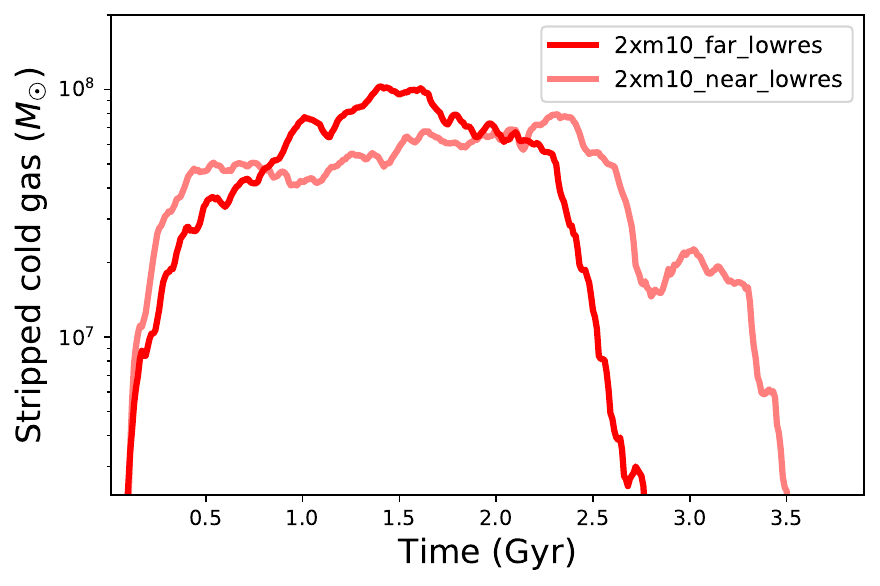}
\caption{{Time evolution of stripped cold gas from one satellite situated at 100 kpc in the case of two different runs of m10 (2xm10\_far\_lowres and 2xm10\_near\_lowres).}}
\label{sig}
\end{figure}

It is important to understand how much the difference between cold gas mass is significant in our study. For that reason, we calculated the time evolution of the cold-stripped gas mass for the m10 satellites which are located at the same position of 100 kpc away from the host in two different simulation runs (Figure \ref{sig}). This will also give a measure of stochasticity in our simulation. We can see the amount of cold gas is not that different until 2.5 Gyr, however after that, there are some differences between the values. Although at late times, satellites are almost gas-deficit, therefore we should take these differences with a pinch of salt. However, at earlier times, the differences between these two runs are less than a factor of two, which implies our runs are not so stochastic, at least until 2.5 Gyr. Therefore, we can take differences in our runs to be significant and independent of stochasticity if they differ by a factor two.       

\normalsize

\label{lastpage}

\end{document}